\documentclass[a4paper,fleqn]{cas-sc}

\usepackage[authoryear,longnamesfirst]{natbib}

\usepackage{algpseudocode}
\usepackage{algorithm}
\usepackage{amsmath}
\usepackage{amssymb}
\usepackage{xcolor}
\def\tsc#1{\csdef{#1}{\textsc{\lowercase{#1}}\xspace}}
\tsc{WGM}
\tsc{QE}
\tsc{EP}
\tsc{PMS}
\tsc{BEC}
\tsc{DE}

\begin{document}
\let\WriteBookmarks\relax
\def\floatpagepagefraction{1}
\def\textpagefraction{.001}
\shorttitle{Parallel dynamic overset grid framework}
\shortauthors{M. Hedayat et~al.}

\title [mode = title]{A parallel dynamic overset grid framework for immersed boundary methods}                      



\author[1]{Mohammadali Hedayat}[type=editor]
\address[1]{J. Mike Walker '66 Department of Mechanical Engineering, Texas A$\&$M University, College Station, USA}

\author[1]{Iman Borazjani}[style=editor,orcid=0000-0001-7940-3168]
\cormark[1]
\cortext[cor1]{Corresponding author}
\ead{iman@tamu.edu}

\begin{abstract}
A parallel dynamic overset framework has been developed for the curvilinear immersed boundary~(overset-CURVIB) method to enable tackling a wide range of challenging flow problems. The dynamic overset grids are used to locally increase the grid resolution near complex immersed bodies, which are handled using a sharp interface immersed boundary method, undergoing large movements as well as arbitrary relative motions. The new framework extends the previous overset-CURVIB method with fixed overset grids and a sequential grid assembly to moving overset grids with an efficient parallel grid assembly. In addition, a new method for the interpolation of variables at the grid boundaries is developed which can drastically decrease the execution time and increase the parallel efficiency of our framework compared to the previous strategy. The moving/rotating overset grids are solved in a non-inertial frame of reference to avoid recalculating the curvilinear metrics of transformation while the background/stationary grids are solved in the inertial frame. The new framework is verified and validated against experimental data, and analytical/benchmark solutions.  In addition, the results of the overset grid are compared with results over a similar single grid. The method is shown to be 2nd order accurate, decrease the computational cost relative to a single grid, and good overall parallel speedup. The grid assembly takes less than 7\% of the total cpu time even at the highest number of cpus tested in this work. The capabilities of our method are demonstrated by simulating the flow past a school of self-propelled aquatic swimmers arranged initially in a diamond pattern.

\end{abstract}




\maketitle

\section{Introduction}

Simulations of unsteady flows with moving/deforming bodies remain challenging due to the constraints and difficulties in mesh generation and boundary condition implementation. Although conforming grids, e.g., arbitrary Lagrangian-Eulerian~(ALE)~\citep{hirt1974arbitrary,donea2017arbitrary,wang2013arbitrary,guardone2011arbitrary,wang2010modeling} have been used for simulating flows around moving bodies, for large displacements ALE methods can result in poor grid quality that decreases the accuracy of  calculations. For large deformations, consequently, remeshing methods~(e.g., global remeshing~\citep{peraire1987adaptive,lohner1989adaptive} or refinement methods~\citep{berger1989local,macneice2000paramesh,kirk2006libmesh,burstedde2011p4est}) are required to rediscretize the whole computational domain or part of it in such a way that the grid is conformal to the structure and the quality of the fluid mesh is preserved as much as possible. However, re-meshing suffers from the loss of accuracy during time evolution due to solution interpolation from the old domain to the new one~\citep{alauzet2018time}. In addition, this method can increase the computational cost of simulations significantly especially for complicated geometries. Furthermore, efficient parallelization of a solver in re-meshing techniques is not straightforward. On the other hand, non-boundary conforming methods, e.g., immersed boundary method~(IBM)~\citep{borazjani2008curvilinear,mori2008implicit,kim2018weak,ma2019hierarchical,wang2019novel}, or fictitious domain method~\citep{baaijens2001fictitious,patankar2000new}, among others, can handle large body deformation but they may decrease the solution's accuracy near the fluid-solid interface due to interpolation errors. In addition, non-boundary conforming methods require high grid resolution in the regions where the boundary movement occurs which can increases the number of grids points relative to ALE methods.

Immersed boundary methods have emerged as a powerful tool to efficiently study complicated real-life flow problems which involve arbitrarily complex bodies/flow domains~\citep{mittal2005immersed,sotiropoulos2014immersed,BHV:Borazjani2014rev}. In these methods, the computational domain is discretized with a single, fixed, non-boundary conforming mesh system which can be curvilinear or Cartesian. Immersed boundary methods have been successfully used for simulations of cardiovascular flows~\citep{peskin1972flow,de2009direct,bavo2016fluid,hedayat2019comparison,gilmanov2018flow,asgharzadeh2019non}, aquatic swimming~\citep{patel2018new,daghooghi2016self}, vortex generation/control~\citep{garg2018sharp,asadi2018scaling,akbarzadeh2019numerical}, etc. Nevertheless, despite many attractive features of the immersed boundary methods, they suffer from a major limitation which raises from the fact that the background grid stays the same and there is no ability for clustering the grid nodes in the boundary layer of moving bodies during a simulation. This limitation makes the application of the immersed boundary method very challenging for flows in which an immersed body undergoes an arbitrary large displacement or rotation, such as aquatic swimmers~\citep{daghooghi2015hydrodynamic}, wind turbines~\citep{li2012dynamic}, or flapping wings~\citep{deng2016dynamic}. In such simulations, the entire background grid should be discretized with a fine grid to resolve the boundary layer near the immersed bodies which increases the computational cost drastically. Although solving Navier-stokes equations in a non-inertial frame of reference can overcome this problem for a single body, the problem still remains for multiple bodies in arbitrary relative motions. 

To address the above issue for moving bodies, a few strategies have been proposed which provide high grid resolution near an immersed body while the grid is coarsend away from the body. Among those are the adaptive mesh refinement~(AMR) and overset grids. Hierarchical AMR technique for Cartesian grids was pioneered by~\citet{berger1984adaptive}. Since then this technique has been applied and developed by many researchers ~\citep{chen1997local,henshaw2008parallel,holst2009overset,angelidis2016unstructured}. Although AMR method is accurate and efficient for steady problems~\citep{alauzet2010high,jones2006validation,michal2012anisotropic} several drawbacks are associated with this method for unsteady flows~\citep{hornung2006managing,alauzet2018time,peng2018approach,angelidis2016unstructured}. The most important problem is the latency between the mesh and flow solution. A few remedies have been proposed in the past few years to overcome this problem. Some work adjust the mesh every $n$ time step and thus the mesh is lagging behind the unsteady solution. However, there is no guarantee that features of interest remain in the refined area in this method~\citep{de1993petrov}. Other strategies such as local adaptive re-meshing~\citep{gruau20053d,compere2008transient} adjust the mesh more frequently. However, errors due to solution interpolation from the old mesh to the new one can generate unquantified errors~\citep{alauzet2018time}. In addition, developing a robust algorithm and data structure for AMR method is usually not straightforward~\citep{angelidis2016unstructured}. Finally, the parallelization of AMR solver with high efficiency is very challenging because the load~(number of grid points) on each computing core is dynamically changing.

Overset or Chimera grids provide an elegant solution for this issue by discretizing a complex flow domain into a set of simpler, overlapping sub-domains which can move relative to each other. The problem of overset grids was first proposed during the 1970s for the solution of the elliptical and hyperbolic partial differential equations for the inviscid shallow-water equations using two-dimensional domains and a non-moving overset grid~\citep{Steger68,starius1977composite,starius1980composite}. \citet{steger1987use} and later \citet{meakin1989unsteady} adopted the idea to tackle more complicated problems of simulating compressible flows around multiple complex geometries. Since then several attempts have been conducted to develop an overset grid framework to handle an arbitrary number of overlapping grids for both compressible and incompressible flows using staggered~\citep{chesshire1990composite,henshaw1994fourth,tang2003overset,vreman2017staggered}, non-staggered grids~\citep{burton2002analysis,burton2005fully}, and hybrid staggered/non-staggered grids~\citep{borazjani2013parallel}. 

In overset grid solvers, the governing equations are solved independently in each sub-domain and the connection between different sub-domains is achieved by interpolating the flow variables at the interface of overlapped domains. The connection between different overlapping domains is established via a grid assembly process. The main tasks in this process are: 1) Hole-cutting; 2) donor search; and 3) variable interpolation. While performing these tasks using a single processor may be trivial, the problem can be very challenging in parallel considering that each grid is partitioned and distributed among several processors. Several grid-assembly packages have been developed in recent years~\citep{zagaris2010toolkit,crabill2018parallel,kenway2017efficient}. All these codes have their advantages and disadvantages. Some of these packages are dedicated assembly codes which provide a general overset grid assembly capability and need a mechanism for integrating with an existing flow-solver~\citep{suhs2002pegasus,noack2009suggar++,alonso2006chimps,sitaraman2010parallel} while others are developed for a specific flow solver and directly added to that solver~\citep{meakin2001object,belk1995automated,wang2000fully,henshaw2002overture,buning2011cartesian,borazjani2013parallel}. Most of dedicated grid assembly packages use out-of-core algorithms to be linked to an existing flow-solver, e.g. through an Input/Output~(I/O) file~\citep{sickles2000time}, which suffer from a high overhead especially for moving overset grids.

In addition, some codes are not implemented in parallel~\citep{borazjani2013parallel}. Because the overset grid assembly should be performed at each time-step as the grids are moving during a simulation, parallel implementation of grid assembly is essential for parallel solvers. However, the scalability of a parallel grid assembly code can be limited due to inherent algorithmic complexity and difficulty of efficiently distributing the computations between available processors. Several strategies are available for handling the grid-assembly task in parallel. Some codes maintain the entire meshes in all processors~\citep{shen2013rans} while others use a partitioning scheme for overset grid assembly which is different from the partitioning used for the flow solver~\citep{miller2014overset,shen2015dynamic}. However, this requires a merge-and-repartition for the entire grid data at each time step of the simulation which can drastically increase the execution time due to memory latency and the algorithm overhead especially in the unsteady flow simulations where these tasks need to be performed at each time step. To overcome the above weaknesses, an algorithm with the capability to handle an already distributed composite grid is required.

Some attempts have been conducted to address the above problems in recent years~\citep{zagaris2010toolkit,roget2014robust,martin2019overset}. \citet{zagaris2010toolkit} developed an in-core parallel grid assembly to tackle the distributed assembly problem. However, the scalability of their method was not satisfying. \citet{roget2014robust} implemented a dynamic load balancing algorithm for PUNDIT~\citep{sitaraman2010parallel}. Although they achieved a very good scalability for a large number of processors, this method is originally developed for unstructured grids and is not suitable for structure grids~\citep{wang2019efficient}. \citet{martin2019overset} developed a grid decomposition method for the the overset grid assembly problem which leads to a scalable computation for very large simulations of moving bodies as well as reduction in memory requirement while the method has some limitations in terms of overlap minimization and optimal donor selection. More recently,~\citet{horne2019massively} developed a massively-parallel overset grid assembly for a PR-DNS of particle-laden turbulent channel flows to simulating large numbers of moving bodies with an exceptional parallel scalability. However, this method is not readily applicable to general curvilinear problems and multi-connected geometries such as cardiovascular system. In addition, the above packages are not easily accessible to a third party and the task of efficiently integrating the existing grid assembly codes to a specific flow solver is problematic. 

In this paper, we developed a new computational framework to extend our previous overset grid code~\citep{borazjani2013parallel} to perform the grid assembly tasks for moving overset grids fully in parallel. The grid assembly is integrated with our sharp-interface CURVIB solver in a general  non-inertial frame of reference with a conservative formulation~\citep{borazjani2008curvilinear,borazjani2013parallel} which provides us the ability to tackle high-resolution, fluid-structure interaction~(FSI) simulations of complex real-life problems. To achieve this goal a number of major algorithmic developments have been presented in this paper compared to the previous work \citep{borazjani2013parallel} which include: 1) developing a new donor search algorithm which enables us to perform the search fully in parallel compared to our previous work which could only run on a single processor; 2) developing a new walking strategy to identify the donor compared to the previous work  which utilized a brute force approach; 3) developing a new parallel interpolation method which can drastically reduce the execution time compared to the previous work; 4) directly integrating the grid assembly kernel to the flow solver to maximize the overlap between computations and data communication compared to the previous work which used an out-of-core method through an I/O file; 5) adding the ability to handle moving overset grids to our CURVIB solver which uses a non-inertial frame of reference for moving girds and an inertial frame of reference for non-moving ones versus our previous work in which all grids were solved either in inertial or non-inertial frame of reference. Our framework has been validated against several experimental and numerical test cases and its versatility is demonstrated by applying it to simulate a challenging FSI simulation of fish swimming in a school.

The paper is organized as follows: The governing equations in general curvilinear coordinates for a non-inertial frame of reference are presented in section~\ref{Numerical-method}. Then, the CURVIB solver is briefly described and the algorithms for parallel overset grid assembly are explained.  Fist, the domain decomposition strategy for flow solver and grid assembly is described, and then the top main functions used for grid assembly, including 1) hole-cutting, 2) donor search, and 3) interpolation are clarified. The optimization of the grid assembly kernel to reduce the overhead using our data packing strategy and eventually the integration of the grid assembly kernel with CURVIB flow solver for moving overset grids are presented. In section~\ref{Numerical-results}, several test cases are presented to show the accuracy and capabilities of our framework. The second-order accuracy of our overset grid approach is verified by comparing it against the analytical solution of Taylor-Green vortex. The rotationally oscillating cylinder is tested to verify the result of our framework for overset grids against a single grid in a rotating frame of reference. Forced inline oscillations of a cylinder is presented as the next test case for our overset framework and the results are compared against both a single grid and experimental results. We also verified the overset results for FSI simulations of falling cylinders in an infinite flow. The results of overset simulations for a single circular cylinder in a free fall under gravitational force are compared against the results of a single grid in a non-inertial frame of reference. Furthermore,  the free fall of multiple circular cylinders are compared to check the ability of our framework in handling simulations involving multiple overset grids with arbitrary motions, and the different scenarios which happen in these simulations. The versatility of our code is demonstrated by the 3D simulations of multiple swimmers in a diamond arrangement. Then, the parallel efficiency of different parts of our framework is tested. Finally, the conclusions and future work are discussed in section~\ref{Conclusions}.

\section{Numerical method} \label{Numerical-method}

Our new method is an extension to our previous method with fixed overset grids~\citep{borazjani2013parallel} which was originally developed to perform grid assembly for non-moving overset grids using a single processor. Our new algorithm can: 1) handle moving overset grids; and 2) perform the grid assembly task fully in parallel for already existing decomposed grid~(grid decomposition is managed by the flow solver). In addition, this grid assembly kernel is integrated into our CIRVIB flow solver which enables us to perform FSI simulations for complicated flow problems using dynamic overset grids. The details of our new grid assembly method and the way it is integrated to our flow solver is discussed in this section.

\subsection{Overset-CURVIB in a non-inertial frame of reference} \label{sec:CURVIB}

The three-dimensional unsteady incompressible continuity and Navier-Stokes equations are the governing equation in the fluid domain and are solved using the curvilinear/immersed boundary~(CURVIB) solver. The CURVIB and fixed overset methods are extensively described and validated in our previous works~\citep{borazjani2013parallel,ge2007numerical,borazjani2008curvilinear,borazjani2013fluid}. Thus, just a brief overview is presented here. 
A fully-curvilinear formulation based on the hybrid staggered/non-staggered approach~\citep{ge2007numerical} is used which eliminate the need for evaluation of the Christoffel symbols. A sharp-interface immersed boundary method is used to handle the 3D, arbitrary complex moving bodies inside the curvilinear background domain which utilizes an efficient ray-tracing algorithm for immersed/fluid node classification \citep{borazjani2008curvilinear}. The boundary conditions are reconstructed on the fluid nodes in the immediate vicinity of the immersed bodies along the normal to the body surface~\citep{gilmanov2005hybrid}. The solver has been shown to be second-order accurate~\citep{borazjani2013parallel,asgharzadeh2017newton}. 

\begin{figure}
	\centering
	\includegraphics[width=0.8\textwidth]{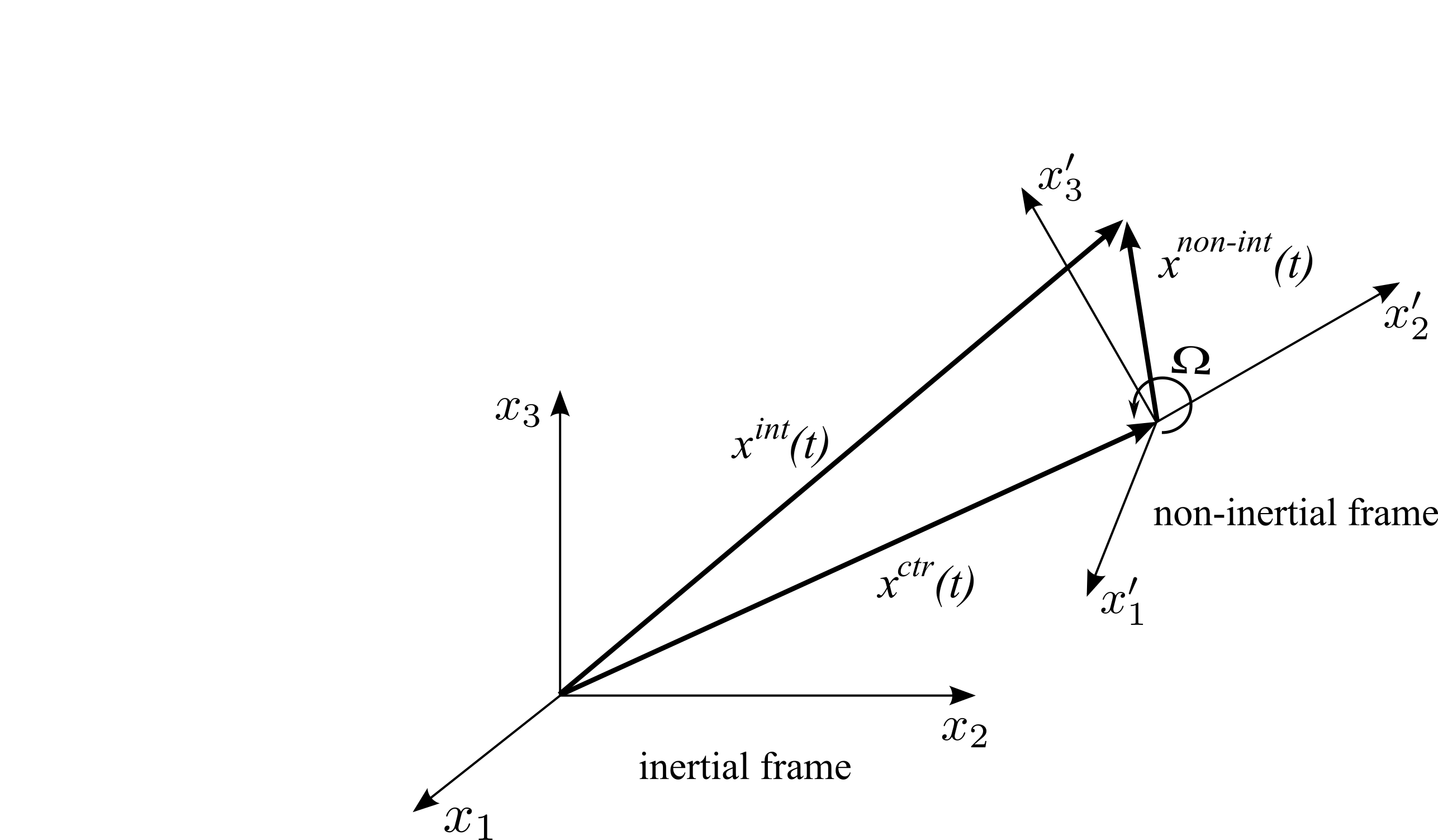}
	\caption{schematic position and orientation of non-inertial frame relative inertial frame where $x^{int}$, $x^{non-int}$ and $x^{ctr}$ are the coordinate vectors in inertial frame, non-inertial frame and origin of non-inertial frame in inertial coordinate.} 
	\label{Fig:non-inerial}
\end{figure}

In this study, a conservative form of Navier-stokes equations in a non-inertial frame of reference for a curvilinear coordinate is employed which was previously developed by~\citet{borazjani2013parallel} based on the work by~\citet{vinokur1989analysis} and \citet{beddhu1996strong}. Fig.~\ref{Fig:non-inerial} illustrates the position and orientation of inertial and non-inertial coordinates relative to each other. Using a general arbitrarily moving non-inertial frame of reference allows us to enhance the versatility and efficiency of our numerical framework for problem involving rigid body motion of an immersed body. Furthermore, using a non-inertial frame of reference for dynamic overset grids enables us to avoid recomputing the metrics of curvilinear transformation at each iteration of the momentum solver where the grid position and orientation changes with respect to the inertial frame of reference which can reduce the computational cost drastically. In addition, when a non-deforming immersed body is present in the fluid domain, using a non-inertial frame can prevent the use of ray-tracing algorithm for the background grid node classification at each time step which also reduces the computational costs especially if a large number of immersed bodies are present in the fluid domain. 

The momentum equations in a non-inertial frame of reference is formulated as follows in tensor notation \citep{borazjani2013parallel}: 

\begin{equation}
    J \frac{\partial}{\partial \xi^r} (U^r) = 0
\label{mom1}
\end{equation}
\begin{equation}
\frac{\partial U^r}{\partial t} =\frac{\xi^r_q}{J} \Big(\frac{\partial u_q}{\partial t}\Big)= \frac{\xi^r_q}{J} \Big(-C_1(u_q)-C_2(w_q)-G_q(p)+\frac{1}{Re} D(u_q)\Big)
\label{mom2}
\end{equation}
where $ \xi^r= \xi^r(x_1, x_2, x_3)$, $r=1,2,3$ are the curvilinear transformation of the Cartesian coordinates $(x_1, x_2, x_3)$ based on the hybrid staggered/non-staggered approach~\citep{ge2007numerical}. $C_1,~C_2,~G,$ and $D$ are the convective, gradient, and viscous operators in curvilinear coordinates
\begin{equation}
    C_1(*)=J \frac{\partial}{\partial \xi^r} \Big[\Big(U^r-V^r\Big) * \Big]
\end{equation}
\begin{equation}
    C_2(*)=J \frac{\partial}{\partial \xi^r} \Big[U^r * \Big]
    \label{conv1}
\end{equation}
\begin{equation}
    G_q(*)=J \frac{\partial}{\partial \xi^r} \Big(\frac{\xi^r_q}{J} * \Big)
\end{equation}
\begin{equation}
    D(*)=J \frac{\partial}{\partial \xi^r} \Big(\frac{g^{rm}}{J} \frac{\partial}{\partial \xi^m} * \Big)
    \label{visc}
\end{equation}
$J$ is the determinant of the Jacobian of the transformation, $J =|\partial (\xi_1 , \xi_2, \xi_3)/ \partial(x_1, x_2, x_3)|$, $\xi^r_q = \frac{\partial{\xi^r}}{\partial{x^q}}$, $g^{rm}$ is the contravariant metric tensor, $g^{rm}= \xi^r_q \xi^m_q$, $U^q$ and $V^q$ are the contravariant velocity components, which are correlated with the Cartesian velocity components as follows:
\begin{equation} \label{flux}
    U^r= u_q \frac{\xi^r_q}{J},~and~ V^r= v_q \frac{\xi^r_q}{J}
\end{equation}
and
\begin{equation} \label{velocity}
    u_q= Q_{qr} u^{int}_r
\end{equation}
\begin{equation}
    v_q= w_{q}+u^{ctr}_{q}
\end{equation}
\begin{equation}
    w_q= \epsilon_{qlm} \Omega_l X^{int}_m
\end{equation}
$u^{ctr}_q= u^{ctr}_q(t)$ and $\Omega_q = \Omega_q(t)$ are the transnational and rotational velocity of the non-inertial frame, respectively, relative to the inertial frame. $Q_{qr}, (q,r = 1,2,3)$ is the orthogonal rotation tensor that rotates the non-inertial frame to the inertial frame orientation. $X^{int}_q$ is component of the position vectors in the inertial reference frame~(for more detail readers can refer to~\citet{borazjani2013parallel}).

The above governing equations are advanced in time using a fractional step method on curvilinear grids \citep{ge2007numerical,borazjani2013parallel}. The momentum equations~(Eqs.~\ref{mom1} and \ref{mom2}) are discretized in time in a fully implicit manner using a second-order backward difference scheme (Italic variables are scalar while the Boldface variables are vectors):
\begin{equation}
   \frac{3\boldsymbol{U}^{(*)}-4\boldsymbol{U}^{(n)}+\boldsymbol{U}^{(n-1)}}{2\Delta t} = RHS(\boldsymbol{U}^{(*)},\boldsymbol{u}^{(*)},p^{(n)})
   \label{rhs}
\end{equation}
where $\boldsymbol{U}$, $\boldsymbol{u}$, and $p$ are the contravariant velocity, Cartesian velocity and pressure, respectively. $n$ denotes the time level and $RHS$ is the right hand side of Eq.~\ref{mom2}. Eq.~\ref{rhs} is solved implicitly using a Newton-Krylov method to obtain the intermediate fluxes $U^{(*)}$. These steps are followed by solving Poisson equation for the pressure correction which is solved using flexible GMRES with multigrid as a preconditioner to obtain divergence-free solution~\citep{ge2007numerical}. The solver is fully parallelized using MPI and PETSc libraries \citep{petsc-user-ref}.

\subsection{Grid assembly for moving overset grids} \label{sec:GridAssembly}

The problem of overset grids refers to the use of multiple disconnected grids that are arbitrarily overlapped to discretize a complex flow domain. The whole domain is partitioned and distributed to every available processor in a way that each processor has a portion of the mesh form all blocks. 
Figure~\ref{Fig:partition} shows the schematic of an arbitrary overset grid with three blocks~(sub-grids) $bi=1$ to $bi=3$ and the distribution of each grid on different processors in our framework at a given time instant. To solve the governing equations on each overset grid, boundary conditions on the nodes at the interface of each block~(e.g., $\Gamma_0$,..., $\Gamma_3$) need to be interpolated from another grid. If a block is enclosed by another one~(e.g., in Fig.~\ref{Fig:partition} $bi=1$ is completely inside $bi=2$), some nodes form the outer block~(here, $bi=2$) in the overlapping region are blanked out to transfer the information from the inner block to the outer block~($\Gamma_4$) by interpolating the solution from the inner domain to several grid points inside the interface of the blanked region. The interpolation on a layer of nodes inside the blanked region, called the buffer layer, is needed to maintain similar discretization stencil on the fluid node in the immediate vicinity of the blanked region as can be observed from Fig.~\ref{Fig:hole-cutting}. The nodes at the interface and/or the blanked region on which the interpolation occurs are known as the query points.

To construct the boundary conditions on the query points, the flow variables are interpolated from the solution of source points known as donors~(from another grid), which may lie in any partition~(each grid can be decomposed in different partitions; in this work, a one-to-one correspondence is present between processors and partitions, e.g. see  Fig.~\ref{Fig:partition}) of that grid. For example, the interface for block $bi=0$, i.e., $\Gamma_0$ needs to be interpolated from $bi=1$ and $bi=2$ grids while the interfaces of $bi=1$, i.e. $\Gamma_1$, needs to be interpolated from $bi=2$ and $bi=0$ grids. Finally, the interface for block $bi=2$, i.e. $\Gamma_2$ needs to be interpolated from block $bi=0$. While performing these operations using a single processor seems trivial, the challenge arises when these tasks are performed in a distributed parallel environments in which data exchange between different domains and different processors/partitions is necessary at each time step.

\begin{figure}
	\centering
	\includegraphics[width=1.\textwidth]{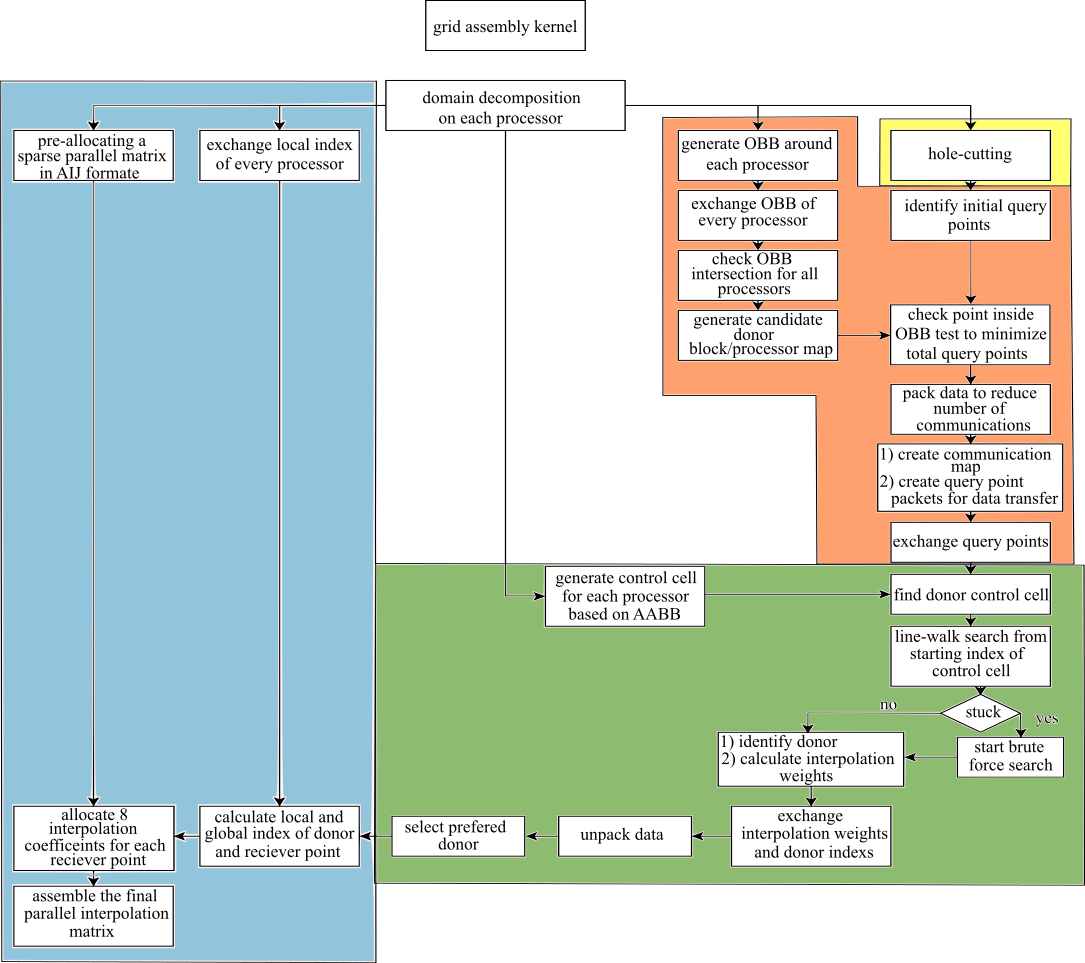}
	\caption{The flow-chart for parallel grid assembly algorithm presented in the paper and forming interpolation matrix} 
	\label{Fig:search-flowchart}
\end{figure}

To achieve a reasonable parallel performance, a new parallel algorithm is developed and implemented which is outlined in Fig.~\ref{Fig:search-flowchart} and will be explained in details in this section. As can be seen in Fig.~\ref{Fig:search-flowchart}, the main steps of our grid assembly method are: 1) domain decomposition which is partitioning and distributing the computational domains to different processors~(section~\ref{decompose}); 2) hole-cutting~(yellow box) which is identifying the grid points that are inside of an immersed body or another grid that need to be blanked out (section~\ref{hole-cut}); 3) query point identification~(orange box) which is performed to identify the points on which the solution needs to be interpolated from another grids/partition (section~\ref{query}) as well as identifying the communication map between processors/partitions and eventually transferring data between different processors (section~\ref{Data-packing}); 4) donor search and donor selection~(green box and section~\ref{donor-search}); and 5) forming the interpolation matrix~(blue box) which is a parallel matrix assembled using the interpolation coefficients obtained during the donor search to interpolate the variables~(velocities) form the donor points to the query points (section~\ref{interpolation}). The detail of each part is provides in sections below.

\subsubsection{Domain decomposition strategy} \label{decompose}

\begin{figure}
	\centering
	\includegraphics[width=1\textwidth]{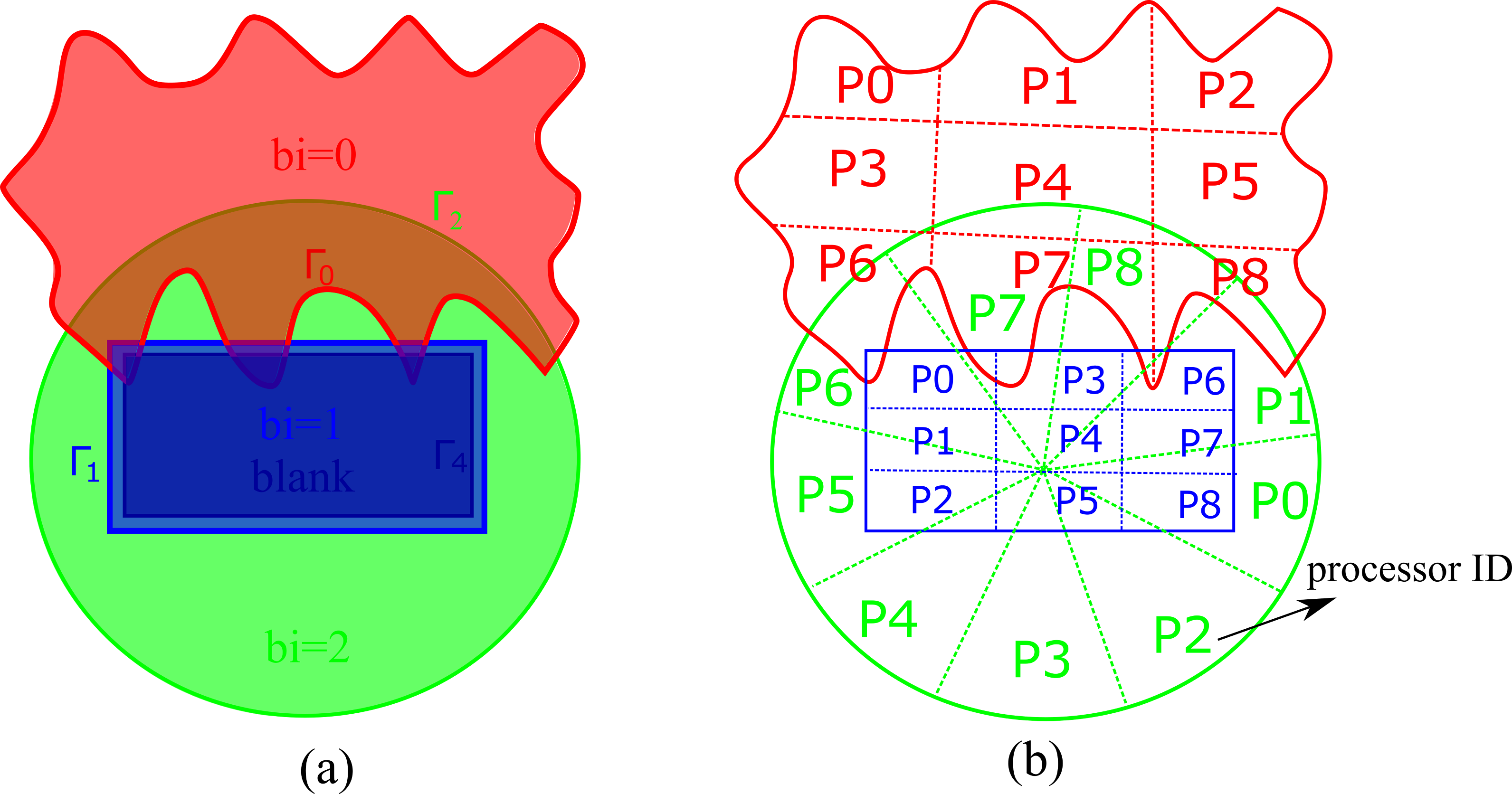}
	\caption{Schematic overset Domain decomposition and domain distribution in different processors for using 9 processors~(p). Every block is distributed over all available processors.} 
	\label{Fig:partition}
\end{figure}

Several factors play a role in the parallel performance of an overset grid solver in terms of both
memory and run-time including 1) the scalability of flow solver, 2) the scalability of the grid assembly method, and 3) the communication between flow and grid assembly solvers. These tasks need to be performed at every time step for a simulation involving moving overset grids in a parallel environment. In this work, message passing interface~(MPI) is used for interprocessor communication required for the overset grid assembly. To reach an acceptable scalability in the flow solver, the workload and, consequently, grid points should be evenly partitioned and distributed among all available processors such that every processor will be involved in solving the flow during the time that the flow solver is running using either implicit and explicit overset coupling. To achieve this goal, the mesh in every block is distributed to all available processors~(e.g. mesh= $[block^{rank= \{0...m\}}_1~...~block^{rank= \{0...m\}}_n$] where $m+1$ = number of processors, and $n$= number of blocks) as illustrated in Fig.~\ref{Fig:partition}. 

To understand the effect of this decomposition on the grid-assembly method, it worth to know that the most time-consuming parts of the grid assembly are hole-cutting and the overhead associated with grid assembly (due to communication/data transfer). This decomposition can increase the performance of the hole-cutting process which works based on the parallel ray-tracing algorithm presented in~\citet{borazjani2008curvilinear} as every processor can separately do the hole-cutting within its partition of each domain. However, it is easy to see that this type of decomposition can drastically increase the number of communications needed between different processors from different blocks in the process of domain assembly (in the worst-case scenario the number communications can reach to $C(n,2) \times P(m,2)$, where $C$ and $P$ are the combination and permutations in Algebra). Two remedies are considered in our framework to treat this problem which can result in decreasing the total number of communication, and, consequently, reducing the total overhead associated with parallelism as well as increasing the overlap between communication and donor search computations. These steps are 1) data packing, and 2) non-blocking data transfer which are explained later in the section~\ref{Data-packing}.

\subsubsection{Hole-cutting}\label{hole-cut}
\begin{figure}
	\centering
	\includegraphics[width=.8\textwidth]{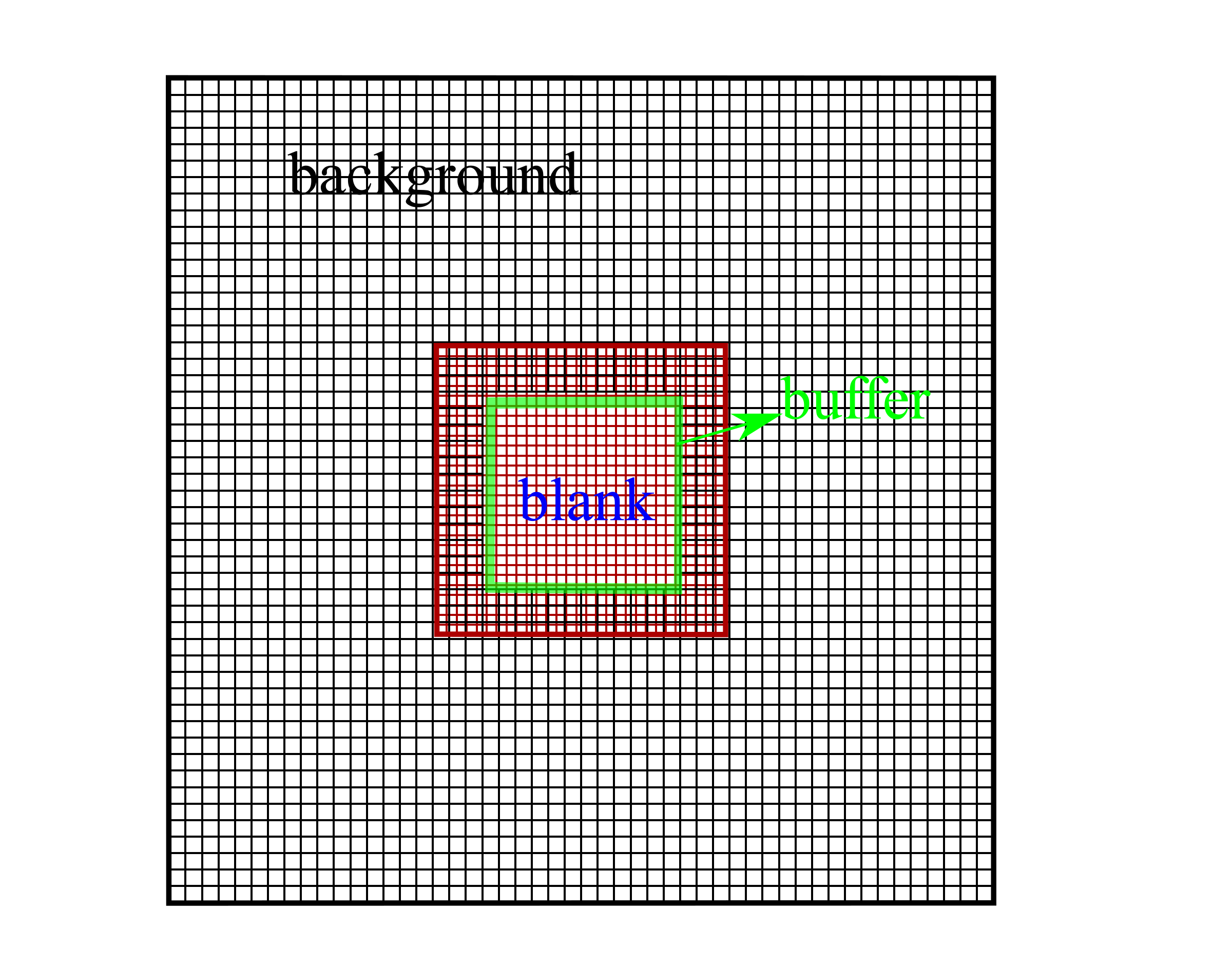}
	\caption{Overset grid layout for the Taylor-Green vortex. The inner domain and blanking region boundaries are shown by thick red and green lines, respectively.} 
	\label{Fig:hole-cutting}
\end{figure}

The first step of the overset grid assembly framework is to identify the blanked (hole) points in the fluid domain. Hole points~(blanked regions) are the points that will be eliminated from the domain, i.e., the flow will not be solved  on these points but interpolated from the inner domain, to transfer information from the inner (typically higher-resolution) domain to the outer one (Fig.~\ref{Fig:hole-cutting}). 

Currently two types of hole-cutting algorithm are available: 1) explicit hole-cutting method~\citep{Overflow,petersson1999algorithm,borazjani2013parallel} in which the user specifies the hole points through the inputs to the algorithm, e.g., user-defined surfaces which are needed for utilizing the ray-tracing algorithm, or 2) implicit hole-cutting methods~\citep{lee2002high,lee2003implicit,liao2007multigrid,hu2019parallel} in which no user-defined input other than the flow solver's boundary conditions are needed. The implicit hole cutting methods work based on an iterative approach for comparing volume grid information and flow boundary conditions to find the best resolution grid. Although implicit hole-cutting methods remove the user-defined inputs to the code, it has approximately an order of magnitude higher computations compared to the explicit one. Therefore, explicit hole-cutting can be more suitable for dynamic overset grids where hole-cutting should be performed at every time step of the flow solver as well as for all strong-coupling iterations within fluid-structure interaction problems \citep{borazjani2008curvilinear}. Hence, in this work, an efficient ray-tracing algorithm similar to the one used for identifying grid nodes located within an immersed boundary in the CURVIB method~\citep{borazjani2013parallel,borazjani2008curvilinear} is used to perform the hole-cutting in the overlapping regions based on the user-defined surface provided to the code as an input.

\subsubsection{Identification of query points} \label{query}

The next step after hole-cutting is to identify the query points on which the variables are needed to be interpolated from another domain. The functions involved in identifying the query points are briefly explained below:

\begin{enumerate}

\item Generate the list of potential query-points: a list of potential query points is formed on every processor, which include the boundary of blank region~(buffer layer in Fig.~\ref{Fig:hole-cutting}) as well as the boundaries of each moving overset grid (points on $i=0,I_{max}$, $j=0,J_{max}$, and $k=0,K_{max}$ where $i$, $j$, $k$ are the grid numbering in curvilinear $\xi^1$, $\xi^2$, and $\xi^3$ directions, respectively).

\item Generate an oriented bounding-box around each processor: To facilitate faster donor search and decreasing the communication time among different processors when dealing with a distributed parallel environment, minimizing the number of query points is essential. Therefore, an oriented bounding box~(OBB) which approximately provides the optimal minimum bounding box is generated around the portion of the distributed grid in each processor as follows:
\begin{equation}
    \textrm{OBB} = \Big\{ctr+ \sum_{n=1}^{3} A_n x_n ~ \big| ~|x_n|<|a_n|, n= \textrm{[1,2,3] for ${\rm I\!R^3}$ space} \Big\}
\end{equation}
and 
where $ctr$ is the center of the grid points, $A_i$ are right-handed orthonormal axes which are calculated as eigenvectors of the covariance matrix of the ghost points (x[min, max], y[min, max], z[min, max]) in each processor, and $a_n$ is the dimension of OBB in $A_i$ direction. Figure~\ref{Fig:OBB} compares the axis-aligned bounding box~(AABB), which will be discussed in section~\ref{donor-search}, and OBB around a sample swimmer. As can be seen in Fig.~\ref{Fig:OBB}, an OBB can provide a tighter bounding box compared to AABB which minimizes the search space for identifying the query points.

\begin{figure}
	\centering
	\includegraphics[width=0.5\textwidth]{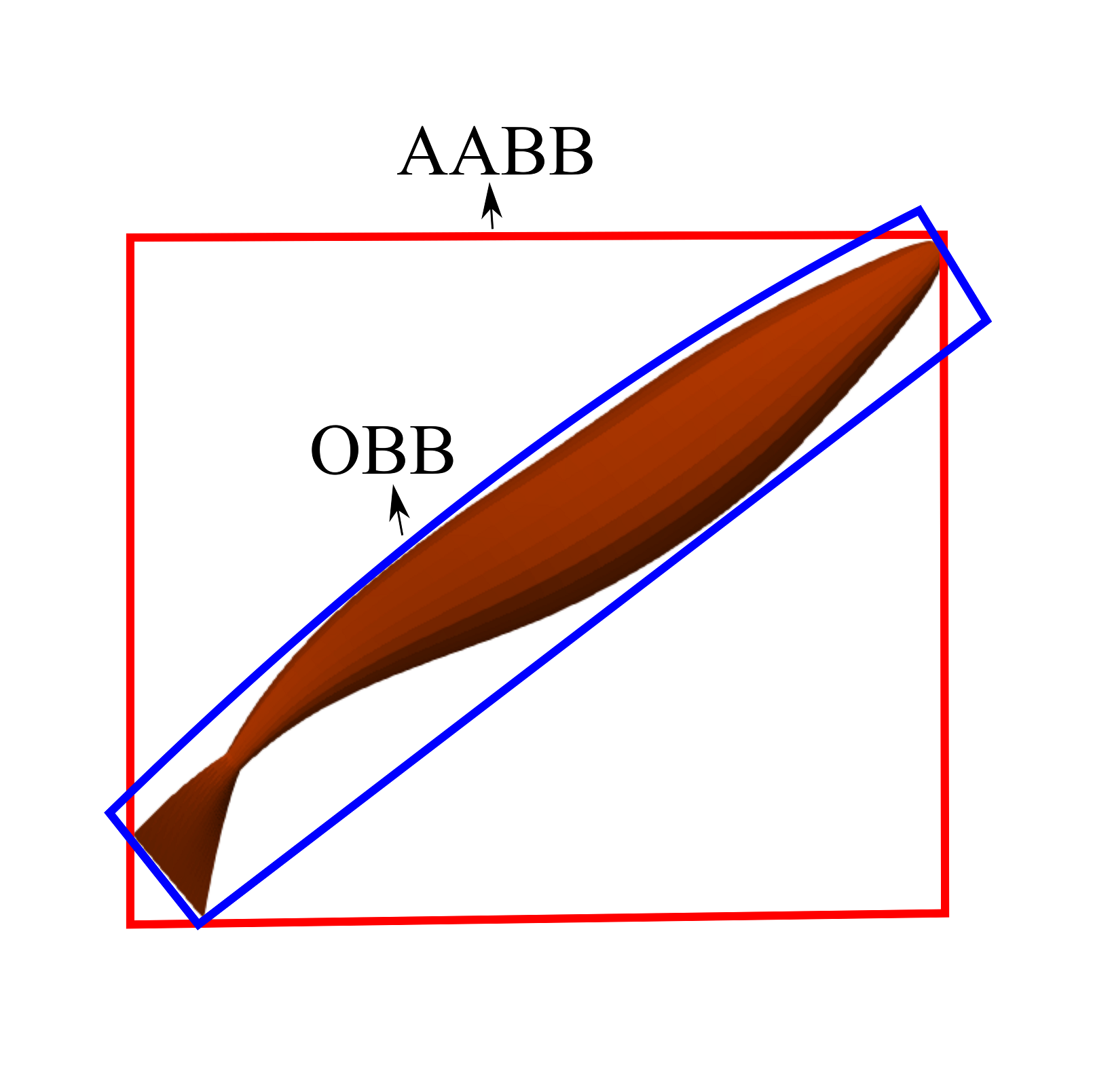}
	\caption{Comparison between an OBB around a swimmer and an AABB which clearly shows OBB provides a much tighter bounding-box around the an object.} 
	\label{Fig:OBB}
\end{figure} 

\item Broadcast the information of OBB of each processor: Since each domain is distributed to all available processors, every block has its own OBB on each processor~(total number of OBBs = number of blocks $\times$ number of processors). The information of the bounding-boxes is then shared between all the processors. Thus, each processor has access to the information of the bounding-box of every other processor in different blocks.

\item Check OBB intersections for all processors: After having all the bounding-box information, a test will be performed to identify the possible intersection of each processor with any other processor. A geometric separation test, explained in~\citet{schneider2003geometric}, is performed to identify the potential intersection of bounding-boxes. Due to our domain decomposition strategy, a total number of $C(n,2) \times P(m,2)$, where $m$ and $n$ are number of processors and blocks, respectively, tests should be performed to find the intersection of different processors.

\item Generate the final list of query points from one processor to the other: After the OBB intersection tests for all the processors, if two processors have intersection with each other the query points should be sent from one processor to the other and vice versa. However, not all the potential query points formed in step $1$ need to be sent to the other processor since none of the potential query points identified in step $1$ may lie inside the other processor's OBB~(even if OBB of two processors can intersects). Thus, to further minimize the number of query points, a point inside OBB test~\citep{schneider2003geometric} is performed~(to check if the potential query point from one processor lies inside the OBB of another processor) to form the final query points list in one processor that needs to be transferred to and be interpolated from another processor. This step helps to reduce the interprocessor communications in the grid assembly method.

\item Data exchange: Following the above steps, the final communication map between all the processors is generated~(if the number of query points in the final list from the previous step is not zero then a communication should be performed otherwise no communication is needed). To transfer data, transfer packets which consist of a list of coordinates of all potential receiver points that need to be interpolated are generated. Then, these packets are exchanged between all the processors based on the final communication map.

\end{enumerate}

After a successful data transfer, every processor will have a list of points for which it needs to perform a donor search. Algorithm~\ref{query-alg} summarizes the query point identification and transfer in our framework.

\begin{algorithm}
  \caption{Algorithm for identifying the query points}
    \label{query-alg}
    \begin{algorithmic}
    \State NP=Number of processors, NB=number of blocks, rank= processor's ID, NQ= number of query points
    \For  {$bi=0$ to $bi=NB$}
    \For  {$P=0$ to $P=NP$}
    \For  {$sb=0$ to $sb=NB$}
    \If{ (sb!=bi $\&$ rank!=P $\&$ (OBB(rank) intersects with OBB(P)))}
    \State potential query points: identify the boundary points in rank
    \State run point~(from rank) inside OBB~(of processor $P$) test
    \If{(point~(from rank) inside OBB~(of processor $P$))}
    \State add point to  query points
    \EndIf
    \State NQ= calculate number of query points
    \State generate transfer packet (send-packet$_\textrm{rank}$[P][QP])
    \State copy query points to send-packet$_\textrm{rank}$[P][QP]
    \State transfer send-packet$_\textrm{rank}$[P][QP]
    \EndIf
    \EndFor 
    \EndFor 
    \EndFor
  \end{algorithmic}
\end{algorithm}

\subsubsection{Data packing strategy} \label{Data-packing}

The schematic of the data packing strategy implemented in this work is outlined in Algorithm~\ref{data-pack}. The data is packed in a way that if the OBB of each two random processors intersects~(section \ref{query}), the data (here, Cartesian coordinates of the receptors) is appended to the transfer buffer regardless of their block number. For the case presented in Fig.~\ref{Fig:partition}, for example, to interpolate on the boundary interface of processor $P=0$ from block $bi=1$ the data should be sent to processor $P=6$~(to be interpolated from $bi=0$ and $bi=2$). Without packing the data, this process should be performed separately, i.e., the information will be sent from Processor $P=0$ to Processor $P=6$ to do the donor search~(section~\ref{donor-search}) for block $bi=0$ and then sent again to do the same process for block $bi=2$. However, the communication from processor $P=0$ to $P=6$ will be performed one time by packing the data for both blocks ($bi$=0, 2). Such data packing helps to reduce the maximum number of the communications to $n \times P(m,2)$ instead of $C(n,2) \times P(m,2)$ for the worst-case scenario which can reduce the overhead related to parallelism (e.g. communications in MPI transfer). In addition, by using a non-blocking communication in data transfer between processors, it allows the algorithm to overlap some of the computations regarding the donor search with communication. In addition, to increase the performance of our framework, the grid assembly is directly linked to the flow solver rather than using any I/O file to exchange the information~(will be explained in section~\ref{interpolation}) which are needed for velocity interpolation on the query points from grid assembly to the flow solver. 

\begin{algorithm}
  \caption{Algorithm for packing data for intercommunication data transfer}
    \label{data-pack}
  \begin{algorithmic}
    \State NP=Number of processors, QP=Number of query points, NB=number of blocks
    \State exchange processors' oriented bounding box
    \State create communication map 
    \For  {$bi=0$ to $bi=NB$}
    \For  {$P=0$ to $P=NP$}
    \For  {$sb=0$ to $sb=NB$}
    \If{ (sb!=bi $\&$ rank!=P $\&$ (OBB(rank) intersects with OBB(P)))} 
    \State append data to send-packet$_\textrm{rank}$[P][QP]
    \EndIf
    \EndFor 
    \EndFor 
    \State send/receive send-packet$_\textrm{rank}$[P][QP]
    \EndFor
  \end{algorithmic}
\end{algorithm}

\subsubsection{Donor search} \label{donor-search}

\begin{figure}
	\centering
	\includegraphics[width=0.5\textwidth]{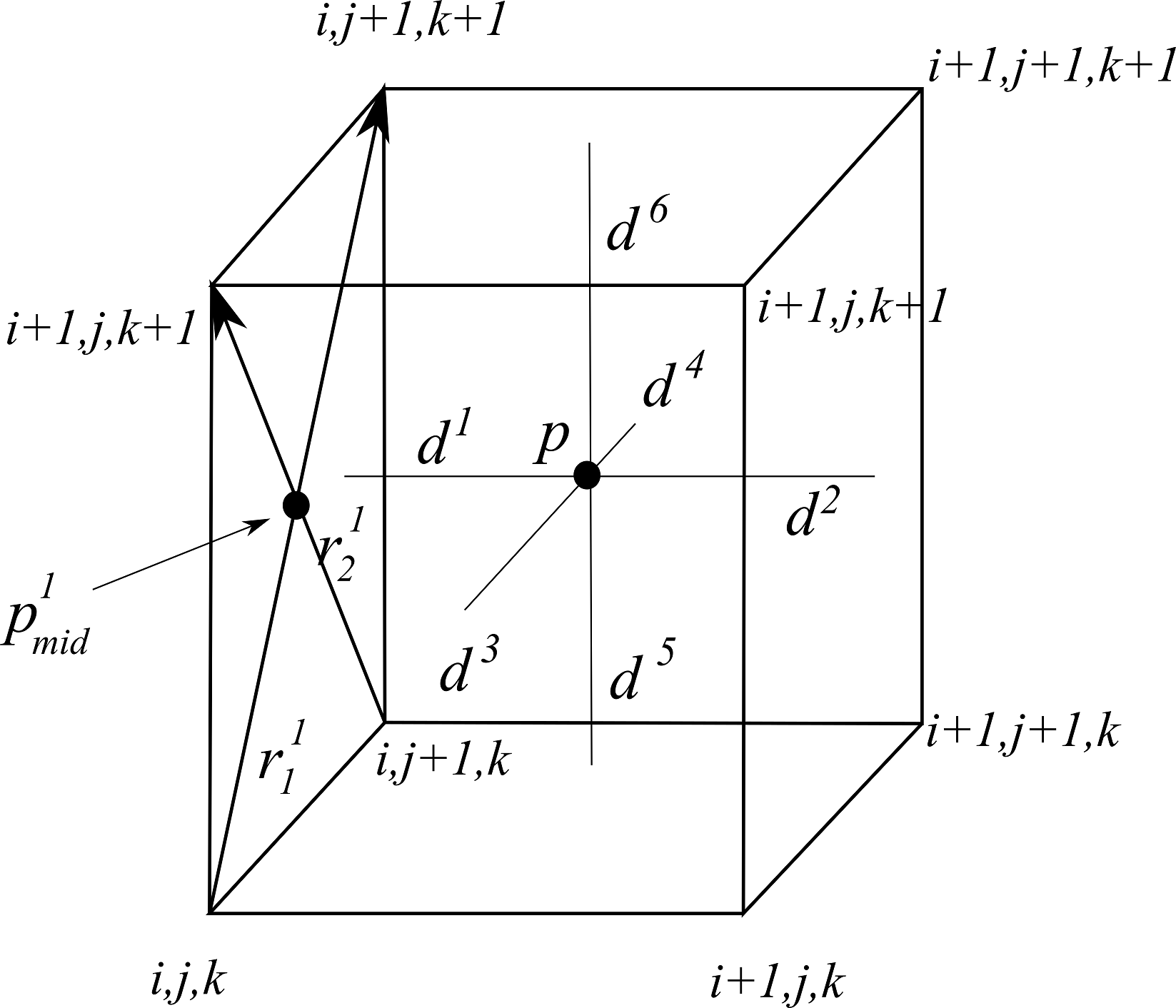}
	\caption{Schematic of search and trilinear interpolation for a point $p$ inside a donor cell. $d^k$ for $k=$ $1$, $6$ shows the distance to the $k^{th}$ surface of the donor cell. The point-in-the-box test can be performed based on $p^1_{mid}$; $r^1_1$, and $r^1_2$ which depict the midpoint and vectors for constructing the inward normal on the $k= 1$ surface of the donor cell. (figure from~\citet{borazjani2013parallel})} 
	\label{Fig:control-cell}
\end{figure}

After receiving the list of query points from all other processors, each processor starts to search for a potential donor. To facilitate the search, localizing the donor is the first step. Hence, an axis-aligned bonding box~(AABB) is generated around the grid partition in each processor to perform a control cell strategy for localization~\citep{borazjani2008curvilinear,borazjani2013parallel} in which an auxiliary grid aligned with Cartesian coordinate is generated around each processor and then divided into several Cartesian boxes, i.e., control cells. The choice of AABB instead of OBB~(Fig.~\ref{Fig:OBB}) is made due to easy implementation and avoiding extra computation required for using OBB. After localization of the donor by finding the proper control cell, the search for the donor cell will start. The donor cell is identified using the point-in-the-box test~\citep{borazjani2013parallel} where the points are the cell corners of each recipient grid while the boxes are the grid cells from the cell centers of the donor grid~(Fig.~\ref{Fig:control-cell}). A point is inside the box if the following inequality is satisfied:
\begin{equation}
    d^{\kappa}=(p-p^\kappa_{mid}). \hat{n}>0
\end{equation}
where $p^\kappa_{mid}$ is the surface center of the $\kappa$th each face of the box and $\hat{n}$ represents the inward unit normal vector to the face which can be computed as $\hat{n}=\frac{r^\kappa_{1} \times r^\kappa_{2}}{r^\kappa_{1} . r^\kappa_{2}}$ in which $r^\kappa_1$ and $r^\kappa_2$ are vectors formed by opposite surface corners of this face as can be seen in Fig.~\ref{Fig:control-cell}. 

In order to increase the speed of the donor search by avoiding a brute force search in each control cell to find a proper donor, a walking search strategy is used. Walking search performs the point-in-the-box test to check if the point is inside the cell and it also utilizes the sign of $d^{\kappa}$ in the above formula to choose the walking direction, e.g. if $d^{\kappa}<0$ it will check the cell in the direction of opposite to the inward normal of $n^\kappa$ and vice versa as outlined below in algorithm~\ref{donor-search_alg}.
\begin{algorithm}
  \caption{Algorithm for donor search}
   \label{donor-search_alg}
  \begin{algorithmic}
    \State find=0
    \State Stuck=0
    \State Locate the control cell $(i_x ,i_y ,i_z )$ where point p is located: 
    \State $i=I(i_x,i_y,i_z)$, $j=J(i_x,i_y,i_z)$, $k=K(i_x,i_y,i_z)$
    \While {(find$<1$)}
    \For {($\kappa=0$ to $\kappa=6$)}
    \If{($d^{1}<0$ $\&$ $d^{2}>0$)}
    \State $i=i-1$
    \ElsIf{($d^{1}>0$ $\&$ $d^{2}<0$)}
    \State $i=i+1$
    \EndIf
    \State do the same for other directions
    \EndFor
    \If{$i=i^{old}~ \& j=j^{old}~\& k=k^{old}$}
    \State Stuck++
    \EndIf
    \If{(Stuck$>0$)}
    \State \textbf{goto} nxtp 
    \EndIf
    \EndWhile
    \State \textbf{nxtp:} search the control cell using brute force search
    
  \end{algorithmic}
\end{algorithm}

\begin{figure}
	\centering
	\includegraphics[width=1\textwidth]{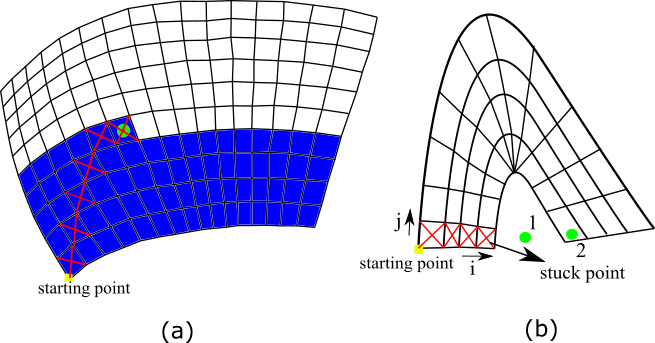}
	\caption{Schematic of search strategies used in this work. a) compares the brute force search (filled blue cells) with walking search~(red cross) b) shows the scenario where the walking search will stuck and thus a brute force search need to be performed.} 
	\label{Fig:search}
\end{figure}

Although the walking search works fine if the receptor point lies inside the boundaries of the donor grid (Fig.~\ref{Fig:search}-a), it will be stuck if the point is outside the boundaries of the donor grid~(Fig.~\ref{Fig:search}-b, point 1) or if the boundaries of the donor grid have a very high curvature inside a control cell  (Fig.~\ref{Fig:search}-b, point $2$). The last scenario  (Fig.~\ref{Fig:search}-b, point $2$) can be prevented by changing the size of the control cell. To overcome this problem in practice, if the location provided by the walking strategy is the same as the previous location, i.e, it is stuck in a cell, the algorithm will break and a brute force search will be performed instead. Using this walking strategy, the cost of donor search reduces from $\mathcal{O}(3)$ in brute force search to $\mathcal{O}(1)$ in the walking strategy. After finding all the donor cells in the donor processor, the communication map between processors generated in section~\ref{query} is reversed and the data buffers~(interpolation coefficients and index~(i,j,k) of the donor) are exchanged back between donor and receiver processors through inter-processor communication.

\subsubsection{Interpolation} \label{interpolation}

Followed by the donor identification, the interpolation coefficients are computed for a trilinear interpolation of flow fields from one domain to the other as follows:
\begin{equation}
\begin{array}{l}
    q_p=a_1 a_2 a_3 q_{i,j,k} + (a_1-1) a_2 a_3 q_{i+1,j,k}+ a_1 (a_2-1) a_3 q_{i,j+1,k}+\\
    (a_1-1) (a_2-1) a_3 q_{i+1,j+1,k}+a_1 a_2 (a_3-1) q_{i,j,k+1}+(a_1-1) a_2 (a_3-1) q_{i+1,j,k+1}\\
    +a_1 (a_2-1) (a_3-1) q_{i,j+1,k+1}+(a_1-1) (a_2-1) (a_3-1) q_{i+1,j+1,k+1}
    \end{array}
\end{equation}
Where $q_p$ is the interpolated flow variable at a query point and $a_i$ are the trilinear interpolation coefficients that are obtained from the distances to the sides (Fig.~\ref{Fig:control-cell}) as follows:
\begin{equation}
    a^1=\frac{d^1}{d^1+d^2}
\end{equation}
\begin{equation}
    a^2=\frac{d^3}{d^3+d^4}
\end{equation}
\begin{equation}
    a^3=\frac{d^5}{d^5+d^6}
\end{equation}
After computing the interpolation coefficients two options are available: 1) directly calculating the interpolated velocities at the donor processor and just return the calculated values~($3$ components of velocity); 2) return the interpolation coefficients~($[a^1,~a^2,~a^3]$) as well as the index~$(i,j,k)$ of the donor and form an interpolation matrix. Using the fist option only $3$ real numbers~($24$ byte for each point) need to be returned while in case of forming an interpolation matrix, $3$ real numbers~($[a^1,~a^2,~a^3]$) as well as $3$ integers~($i,~j,~k$ of donor) should be transferred which will be a total of $36$ bytes of data.  
Although for forming an interpolation matrix more data needs to be transferred, the matrix formation needs to be performed only once before the iterations of the Newton-Krylov method for momentum equation~(Eq.~\ref{rhs}) begins. 
In addition, by using an interpolation matrix and using available toolkits, e.g., PETSc~\citep{petsc-user-ref}, which utilize highly optimized libraries and parallel algorithms for matrix multiplication, the interpolated values are obtained by performing a matrix-vector multiplication and thus will be very robust. Furthermore, the interpolation method will be general and can also be easily used for any flow variable, e.g. scalar concentration, etc. Therefore, the interpolation matrix is used in this work, which has also been more efficient based on our numerical tests.
Compared to our previous method~\citep{borazjani2013parallel} in which the domain connectivity information was calculated using a single processor and then the information were broadcast to all other processors~(which obviously is not efficient and thus not suitable for large scale problems), our current interpolation method is quite faster and more efficient for a large number of grid points.

\begin{figure}
	\centering
	\includegraphics[width=1\textwidth]{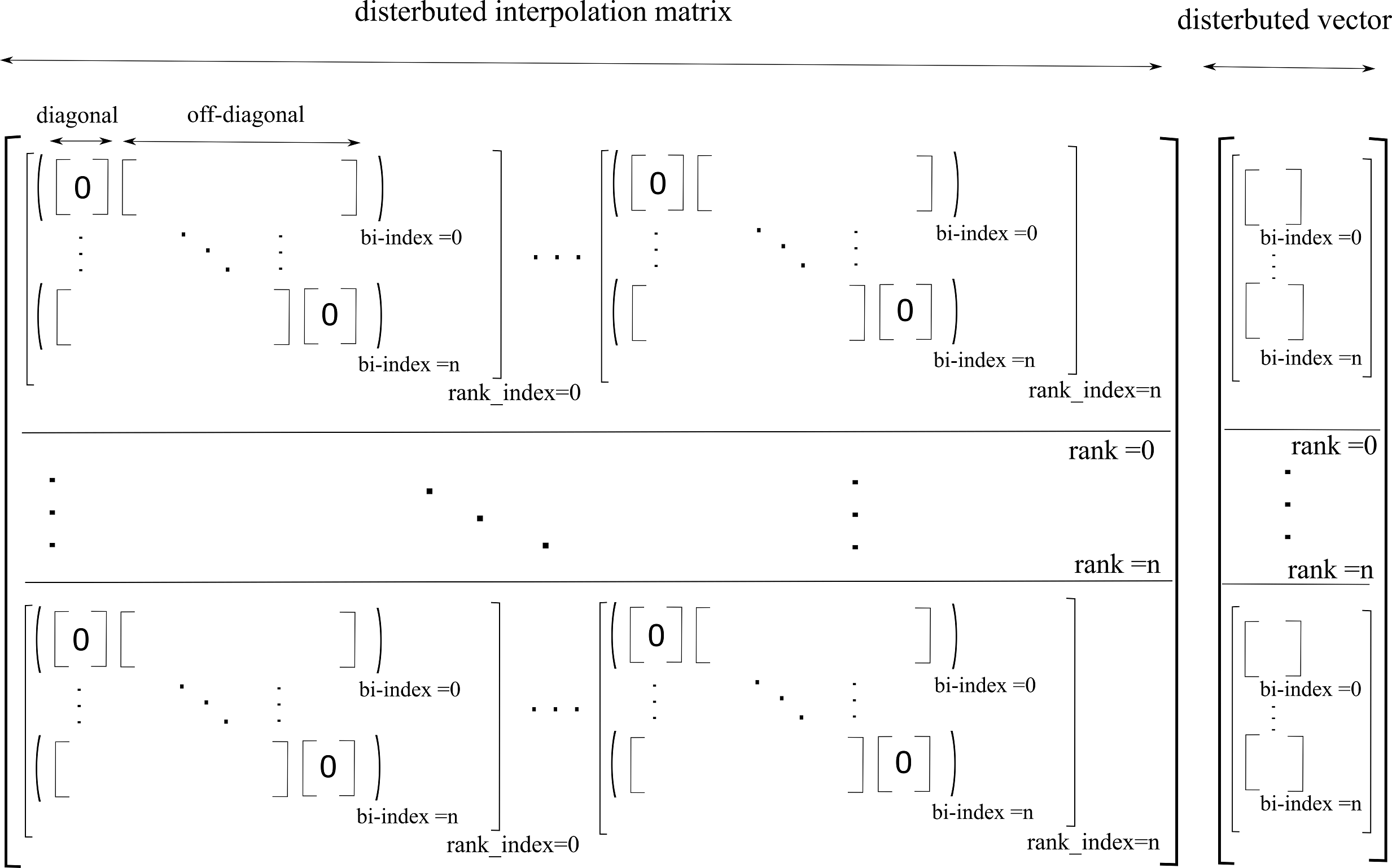}
	\caption{Schematic of assembled parallel interpolation matrix and distributed field vector. Here, rank shows the ID of each processor, rank$\_$index shows index of the processor from which the solution is interpolated, and bi$\_$index shows the block number in each processor.} 
	\label{Fig:interpolation-matrix}
\end{figure}

In this work, PETSc toolkit~\citep{petsc-user-ref} is used for parallel matrix assembly and matrix-vector multiplication. The final interpolation matrix is stored in a compressed sparse row format to minimize memory storage. However, forming a parallel interpolation matrix efficiently is not a very straightforward task. The first step to allocate the interpolation matrix is to define a local and a global index for the points in all domains and processors. Fig.~\ref{Fig:interpolation-matrix} shows the architecture of the allocated matrix and the parallel vector of flow variables. As it can be seen in the parallel distributed vector of the flow variables (Fig.~\ref{Fig:interpolation-matrix}), each processor packs the variables in that processor for different blocks one after the other based on their block number~(bi), e.g. processor zero~($P=0$) packs all the velocity vectors $u_{p=0}=\{u_{bi=0},...,u_{bi=n}\}$. Based on this strategy the local index~($L\_index^{bi}_{P}$) in each processor $P$ for block $bi$ can be defined as follows:
\begin{equation}
L\_index^{bi}_{P}(i,j,k) = \Big(I^0_x+I_x\times I^0_y+ I^0_x\times I^0_y\times I^0_z \Big)^0_P+\\~...
+\Big(i+I^{bi}_x\times j+ I^{bi}_x\times I^{bi}_y\times k \Big)^{bi}_P
\end{equation}
where $(i,j,k)$ are the index of the point and $[I_x,I_y,I_z]$ are dimensions of distributed grid in processor $P$ in $x,y,z$ directions, respectively. Since the donor cell can be in any processor and any block, to be able to define a global index~($G\_index^{bi}_{P}(i,j,k)$), it is necessary for all the processors to know the domain decomposition pattern for every domain. Since we are using a structured grid, the starting and end grid numbers for all curvilinear coordinates in every processor will be enough to define the global index as follows:
\begin{equation}
G\_index^{bi}_{P}(i,j,k) = \sum^{P-1}_{proc=0} \sum^{n}_{block=0} L\_index^{block}_{Proc} + L\_index^{bi}_{P}(i,j,k)
\end{equation}
where $P$ and $n$ are the processor's number and the number of blocks, respectively. After the indexing is done, each processor will form a portion of the interpolation matrix related to its grid point and then the whole matrix will be assembled. The Algorithm~\ref{int_matrix} outlines the process for the parallel interpolation matrix assembly. 

\begin{algorithm}
  \caption{Algorithm for interpolation matrix assembly}
  \label{int_matrix}
  \begin{algorithmic}
    \State get donor-index and interpolation coefficients
    \State $NQ$=Number of query points
    \For  {$i=0$ to $i=NQ$}
    \If{ (donor is available)} 
    \State identify ID of the donor processor
    \State column-index= global index for the corners of the donor cell
    \State row-index= global index for the receptor point in processor rank
    \EndIf
    \EndFor
    \State Creates a sparse parallel interpolation matrix in AIJ format
  \end{algorithmic}
\end{algorithm}

\subsection{Moving overset-CURVIB flow solver}

\begin{figure}
	\centering
	\includegraphics[width=1.\textwidth]{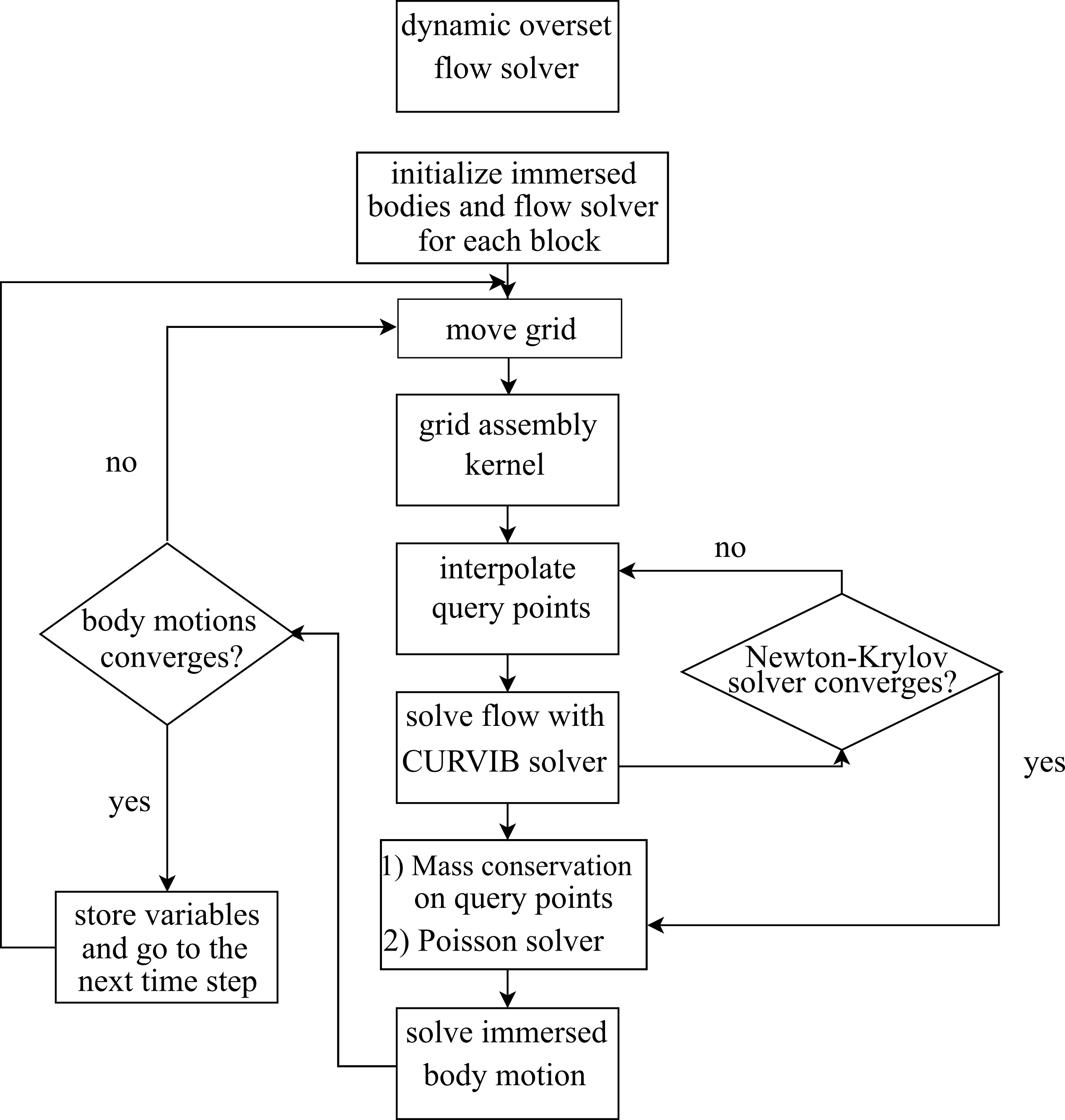}
	\caption{Integrating grid assembly kernel to CURVIB flow solver for FSI simulations} 
	\label{Fig:flow-flowchart}
\end{figure}

The above grid assembly is directly integrated to our CURVIB flow solver. Fig.~\ref{Fig:flow-flowchart} illustrates the flow-chart for the flow solver and how the grid assembly and the interpolation kernels are integrated to our CURVIB flow solver. At the beginning of each simulation, the location of immersed bodies and the flow is initialized. Then for dynamic overset grids, the overset grid over each immersed body is moved based on the motion of that immersed body, i.e., the overset grids are moved with the center of mass of that immeresed bodies. This movement can be either prescribed or calculated based on hydrodynamic forces applied to each immersed body for FSI simulations. After each grid movement, the grid assembly task (section~\ref{sec:GridAssembly}) is performed because the relative position of the grids and, consequently, the donors and interpolation coefficients have changed. After performing the grid assembly and obtaining the query points, donors, and interpolation coefficients, the velocities are interpolated and the fluxes are reconstructed on the query points as will be explained in section~\ref{reconstruction}. Since a non-inertial frame of reference is used for solving the flow in this work, the velocities cannot be interpolated directly from one domain to the other, thus, a transformation from the reference of one domain to the other is needed~(see section~\ref{reconstruction}). Having the fluxes at the interfaces and the buffer layer, the flow is solved using the CURVIB method (section~\ref{sec:CURVIB}). Since an implicit method using a Newton-Krylov solver is used for solving the momentum equation in this work, the interpolation need to be performed in each iteration of the Newton-Krylov solver. After solving the momentum equation, the mass conservation should explicitly be satisfied on the query points~(see section~\ref{mass-conservation}), and then the Poisson equation for the correction step is solved to enforce continuity. Furthermore, for strong-coupling FSI simulations all the above steps should be performed in every sub-iteration of the strong-coupling iterations until the desired convergence for criteria in the structure solver is satisfied \citep{borazjani2008curvilinear}.

\subsubsection{Flux reconstruction and velocity transformation between inertial and non-inertial frames}\label{reconstruction}

Following the interpolation process, the interpolated velocities will be available in each processor. Because a non-inertial frame of reference is used for solving the momentum equations in each block of the domain, however, the interpolated velocities will be in a non-inertial frame and cannot be used directly in another domain since based on Eq.~\ref{velocity_non} the non-inertial velocity is related to inertial velocity by an orthogonal rotation tensor. Using the communication map generated in section~\ref{query}, the donor and receiver blocks are known and the velocities from one domain to the other can be converted using the following formula:
\begin{equation} \label{velocity_non}
    \Big\{u^{non-int}_p\big\}_{(bi=m)}= \Big(Q^{bi=m}_{pr} \Big) \Big(Q^{bi=n}_{qr} \Big)^{-1} \Big\{u^{non-int}_q\big\}_{(bi=n)}
\end{equation}
where $\Big\{u^{non-int}_p\Big\}_{(bi)}$ is the non-inertial velocity in block $bi$ and $Q^{m}_{pr}$ is an orthogonal rotation matrix~(note, $\Big(Q^{n}_{qr} \Big)^{-1}=Q^{n}_{rq}$) which relates coordinates of block $bi=m$ to inertial coordinates. Finally, the flux $U^r$ for each recipient cell surface is obtained by Eq.~\ref{flux} using the scheme discussed in \citet{borazjani2013parallel}.

\subsection{Handling special cases: Overlap of overset boundaries with immersed boundaries or other overset grids} \label{Handling_special_cases}

To be able to apply the above algorithm (Fig.~\ref{Fig:flow-flowchart}) when multiple overset grids or immersed boundaries overlap after grid motion, some special cases should be considered. These considerations are as follows:
\begin{enumerate}
\item Donor selection: In case of multiple overlap grids in the simulation, there may be multiple donors available for a query point. There are several ways to select the donor in these situations. In this study, the query points for each blanked region of the background grid will be interpolated from a specified moving overset which will be provided to the code as an input, whereas the interface of a moving grid inside the background grids is either interpolated from other moving grids or the background grid. To interpolate the interface of a moving grid, the priority is given to the background grid, but it will be interpolated from other available overset grids if a suitable donor does not exist in the background grid, e.g., the donor in the background grid is a blanked node.  

\item Interpolation near solid wall boundaries: Another case happens when the interface of one grid crosses/intersects a solid body or a wall boundary condition. Fig.~\ref{Fig:collision_grid} illustrates this situation where the interface of the red and blue grids are crossing/intersecting immersed bodies. 
Therefore, some of the nodes of the donor cell (Fig.~\ref{Fig:control-cell}) might be inside the immersed body, i.e., a velocity inside the immersed boundary is needed for interpolating onto the interface of the moving grid. Since the flow is inside the solid body is not available, two options are possible: 1) assigning an approximate velocity to solid nodes inside the immersed body based on the velocity of the body; or 2) blanking the region around the immersed body in the other domain (Fig.~\ref{Fig:overset-grid1}). Assigning an approximate velocity to solid nodes will reduce the accuracy of simulations especially in FSI simulations. Hence, the area near the immersed boundary is blanked out and the boundaries of this region is interpolated from the other moving grid (Fig.~\ref{Fig:overset-grid1}). Fig.~\ref{Fig:collision_grid} shows the position of the blanked region of the background grid while Fig.~\ref{Fig:overset-grid1} shows the position of the blanked region of the moving overset grid.

\begin{figure}
	\centering
	\includegraphics[width=0.7\textwidth]{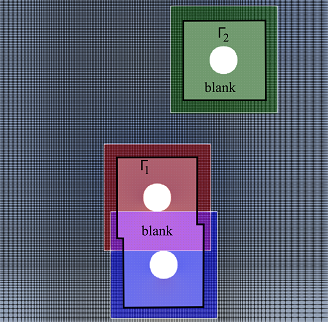}
	\caption{Illustrates the position of overset grids and blanked regions of the background domain for a simulation with bodies in relative motion~(here, circular cylinders) and the way the conservation of mass is satisfied. Here, the blanked region in the red grid intersects with one in the blue grid. In this situation, the mass is conserved over $\Gamma_1$ which is the boundary of the combined blanked region of red and blue grids. However, in the situation without intersection, e.g., $\Gamma_1$ and $\Gamma_2$, the mass is conserved on each blanked region separately.} 
	\label{Fig:collision_grid}
\end{figure} 

\end{enumerate}

\subsubsection{Mass conservation} \label{mass-conservation}

\begin{figure}
	\centering
	\includegraphics[width=.8\textwidth]{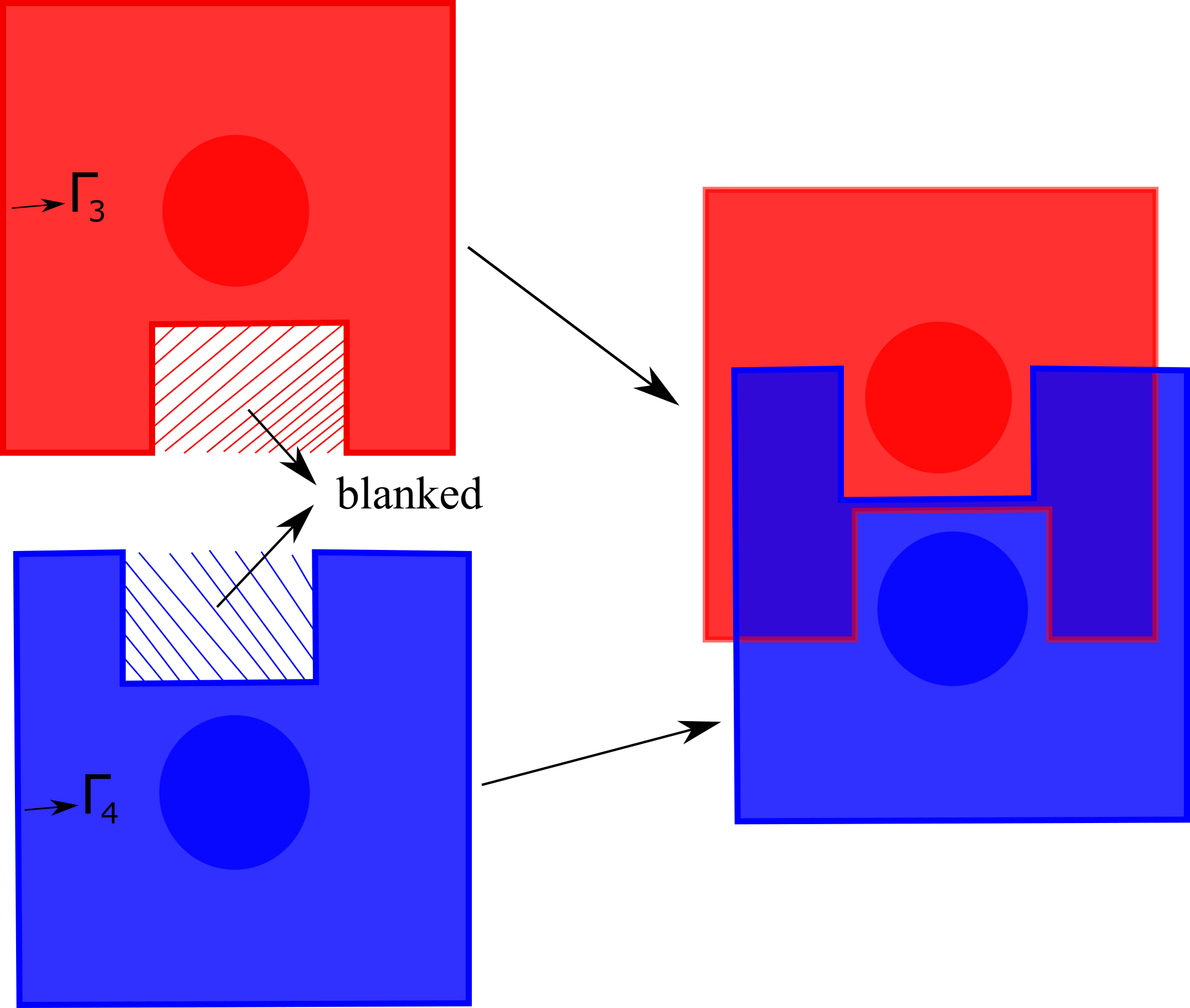}
	\caption{Demonstrates the the strategy chosen for the multiple bodies in relative motion where the interface of overset grid intersects with a solid body. The area around the body is blanked out from the other overset grid, e.g. the area around the res body is blanked out from the blue overset grid, which prevents intersection of interface with solid body. For each interface the mass is conserved separately, e.g. mass conservation is satisfied on $\Gamma_1$ and $\Gamma_2$ separately.} 
	\label{Fig:overset-grid1}
\end{figure}

For incompressible flows, the flux over any closed, non-deforming surface $\Gamma$ within the fluid should be zero:
\begin{equation}
\int_\Gamma \boldsymbol{u} . \hat{\boldsymbol{n}} d\Gamma=0
\end{equation} 
where $\hat{\boldsymbol{n}}$ is the outward normal to the boundary. Therefore, the flux over blanked regions (Fig.~\ref{Fig:collision_grid}) or the interface of overset grids needs to be zero (Fig.~\ref{Fig:overset-grid1}). 
However, the flux on the interfaces are reconstructed using the most recent solution of the donor domain based on the intermediate velocities~($\boldsymbol{u}^{*}$) in the projection method (Eqn.~\ref{rhs}). Since the $\boldsymbol{u}^{*}$ does not satisfy continuity equation and a trilinear interpolation is not a conservative scheme, the global conservation of mass is not satisfied at the overset grid interfaces~(for more details refer to \citet{borazjani2013parallel}). The mass conservation is enforced by adding a correction to the flux, which is calculated by setting the summation of fluxes at the interfaces of each domain and over the blanked region to zero similar to the non-moving overset grids \citep{borazjani2013parallel}. For multiple body collisions~(Fig.~\ref{Fig:collision_grid}), the summation of fluxes at all the blanks together are forced to be zero, e.g., the flux over $\Gamma_1$ and $\Gamma_2$ are forced to be zero in the background grid (Fig.~\ref{Fig:collision_grid}) while the flux over $\Gamma_3$ and $\Gamma_4$ which are the boundaries of overset grids are forced to be zero separately (Fig.~\ref{Fig:overset-grid1}).

\section{Numerical results}\label{Numerical-results}

In this section, we apply the numerical method to simulate several different flows including Taylor-Green vortex, flow past a rotationally oscillating cylinder, forced inline oscillations of a cylinder, free-falling of single and multiple circular cylinders in a fluid, and a school of mackerels in the diamond arrangement. Finally, the parallel efficiency of our framework is presented.

\subsection{Taylor-Green vortex} 

\begin{figure}
	\centering
	\includegraphics[width=1.\textwidth]{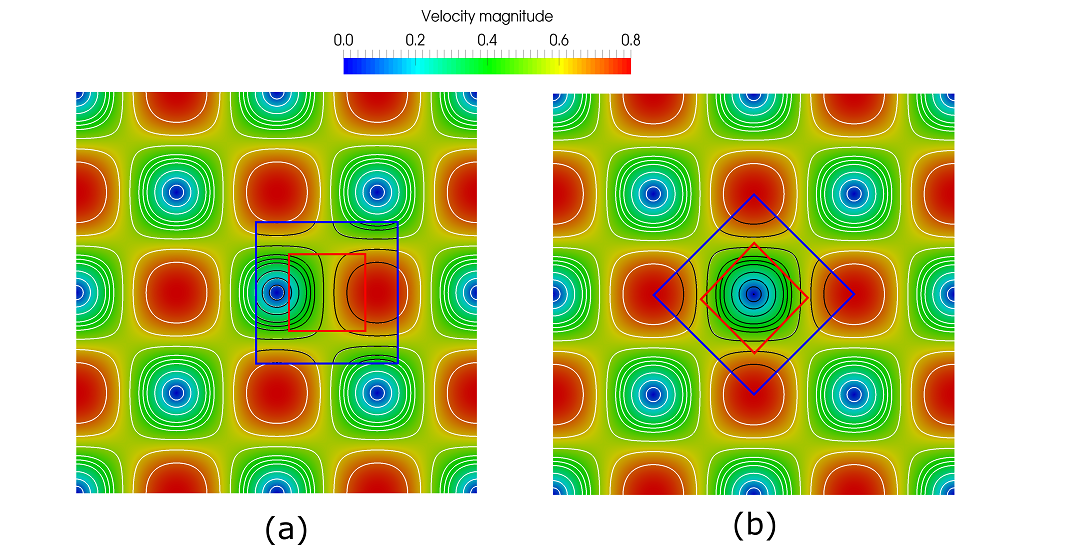}
	\caption{Taylor-Green vortex problem at $t=1$ and $Re=10$: (a) the contour of velocity magnitude for a moving overset with translational speed of $V=PI/4$ b) the contour of velocity magnitude for a rotating overset with rotational speed of $\omega_c=PI/4$. It also compares the streamline contours for both overset grid~(black) and background grid~(white)} 
	\label{Fig:green_taylor}
\end{figure}

The Taylor-Green vortex problem is adopted to investigate the performance and accuracy of the dynamic overset-CURVIB framework. Two-dimensional Taylor-Green vortex is an unsteady flow of a decaying vortex with periodic boundary conditions in two directions~(here, $x$ and $y$) and symmetric in the other direction~(here, $z$). Existence of an exact analytical solution that satisfies the $2D$ incompressible Navier-Stokes equations makes Taylor-Green vortex a suitable benchmark to examine the precision of the computational results. The initial condition is the analytical solution at $t = 0$ in all the domains for all the simulations. The background domain with the size of $2\pi \times 2\pi$ in periodic directions~($x$ and $y$) is discretized uniformly with $201 \times 201$ nodes. An square overset grid with the dimension of $2.2$ centered at the center of coordinates at the initial condition is discretized uniformly with $121 \times 121$ nodes. In addition, a square blank region with the size of $1.5$ on each side is used to blank out the nodes inside the background region. To test the accuracy of our framework for a moving overset grid, two test cases, one for an overset grid with translational movement and the other one with rotational moving are tested. For the translationally moving overset grid, the overset grid is moving with time using a constant translational velocity of $V=\pi/4$ in $x$ direction. In the rotating case, the overset grid is rotating with a constant rotational velocity of $\omega_c= \pi/4$ around $z$ axis. Figure~\ref{Fig:green_taylor} show the contour of velocity and the streamline at a cross-section of the computational domain at t=1 ($\Delta t= 2.5 \times 10^{-4}$) in the simulation with Reynolds number~($Re$) equal to $10$. As can be observed in Fig.~\ref{Fig:green_taylor} the contour of streamlines in the overset domain and the background domain almost exactly match with each other for both rotational and translational moving overset. 

\begin{figure}
	\centering
	\includegraphics[width=0.9\textwidth]{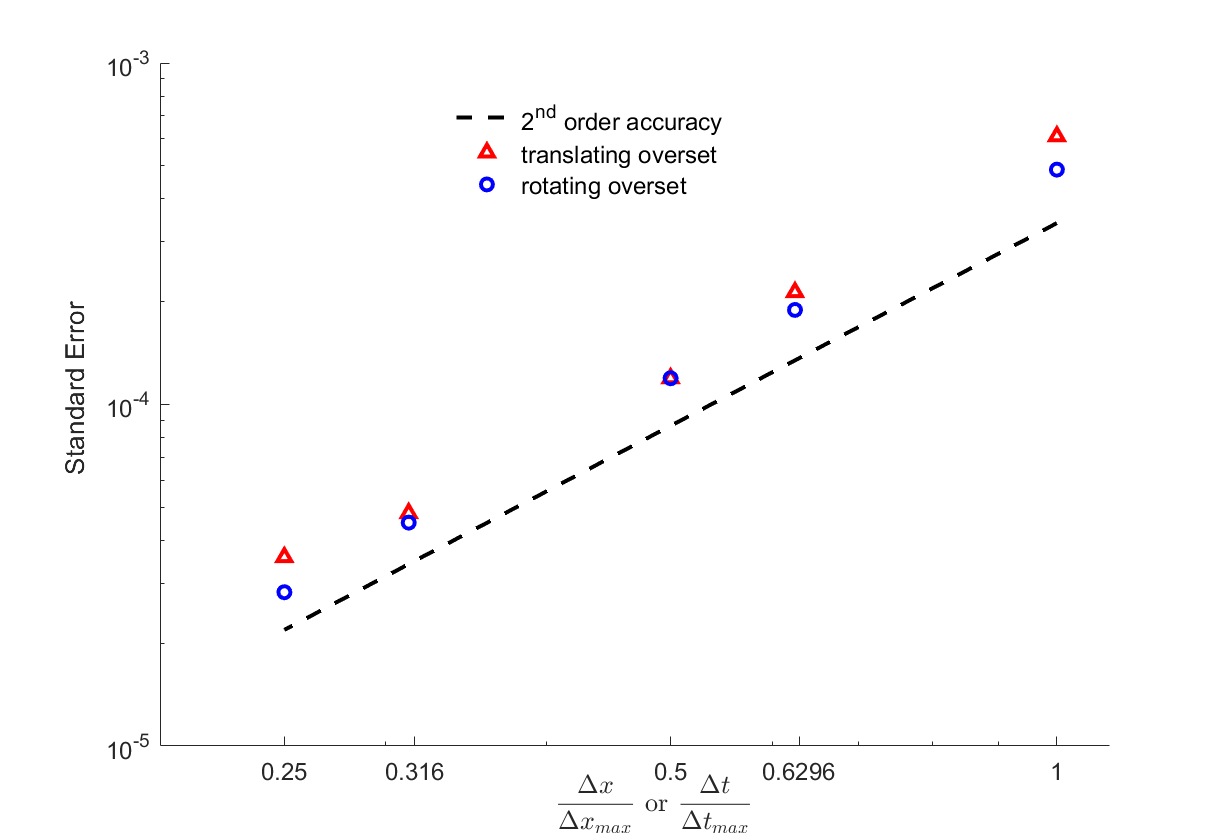}
	\caption{Standard error for Taylor-Green vortex flow as a function of the grid spacing/time-step in log–log scale at $t=1$ and $Re=10$. $CFL$ is kept the same for all test cases, i.e.,$\frac{\Delta t}{\Delta t_{max}}=\frac{\Delta x}{\Delta x_{max}}$. The error shows about $2^{nd}$ order reduction with mesh refinement.} 
	\label{Fig:accuracy}
\end{figure}

The accuracy of the solver in time and space is calculated by computing the error of the numerical results compared to the analytical solution for five simulations with grid size of~(size of larger domain $\times$ size of smaller domain) of $51 \times 31$, $81 \times 49$, $81 \times 49$, $101 \times 61$, $161 \times 98$, $ 201 \times 121$ and time-steps of $\Delta t =$ $1 \times 10^{-2}$, $6.29 \times 10^{-3}$, $5 \times 10^{-3}$, $3.16 \times 10^{-3}$, $2.5 \times 10^{-3}$ for the largest to smallest grids, respectively. The standard error is used to calculate the error in the computational domain as follows:
\begin{equation} 
    \textrm{Standard Error}= \sum_{b=1}^{b=2} \frac{1}{N_x^b N_y^b} \sqrt{\sum_{i=1}^{i=N_x^b} \sum_{i=1}^{i=N_y^b} (u_{(i,j)}^b-u_{(i,j)}^{exact})^2+(v_{(i,j)}^b-v_{(i,j)}^{exact})^2}
\end{equation}
where $N_x^b$ and $N_y^b$ are the number of grid points in $i$ and $j$ direction, respectively, $u_{(i,j)}^b$ and $v_{(i,j)}^b$ are the numerical solutions of velocities on the $(i,j)$ grid point of each sub-grid~($b$) of the overset grid, and $u_{(i,j)}^{exact}$ and $v_{(i,j)}^{exact}$ are the analytical solutions. Figure~\ref{Fig:accuracy} plots the error against grid spacing and time step for both rotational and transitional overset in Taylor-Green vortex problem at $t=1$ and $Re=10$ in log-log scale and demonstrates that the error reduces with about second order accuracy with grid/time refinement.

\clearpage

\subsection{Rotationally oscillating cylinder}

\begin{figure}
	\centering
	\includegraphics[width=1.\textwidth]{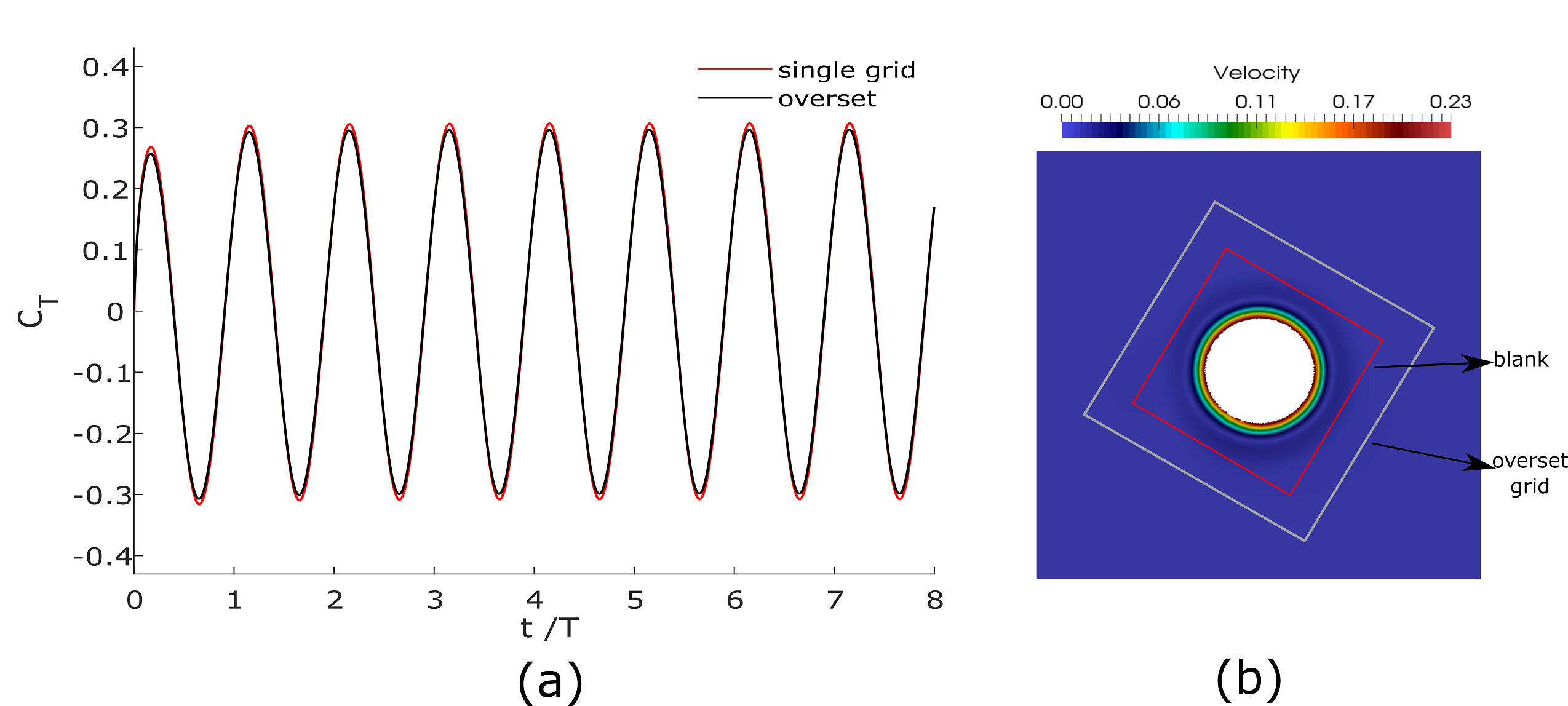}
	\caption{Time histories of the torque coefficient for the flow around a rotationally oscillating cylinder at $Re= 300$.} 
	\label{Fig:rotcylinder}
\end{figure}

To test the ability of overset-CURVIB for rotational immersed bodies, we simulate a rotationally oscillating cylinder in an initially stagnant fluid. The simulations are performed in the inertial frame of reference for the background grid and the non-inertial frame of reference attached to the cylinder for the overset grid. The cylinder is rotating with a rotational motion prescribed by a harmonic oscillation as follows:
\begin{equation} 
\omega_c(t)= A_m sin(2 \pi ft)
\end{equation}
where $\omega_c(t)$, $A_m$, and $f$ are the angular velocity of the cylinder, amplitude, and frequency of the oscillation, respectively. Consequently, the Reynolds number can be defined as $Re=U_m D/\nu$, where $U_m=A_m D/2$, $D$ is the diameter of the cylinder, and $\nu$ is the kinematic viscosity. In this study, all the parameters including the flow parameters, domain size, boundary conditions are chosen similar to~\citep{kim2006immersed}. The size of the background domain is $50D < x_r < 50D$ and $50D < y_r < 50D$. The Reynolds number is defined as $Re=U_m D/\nu$, where $U_m=A_m D/2$, $D$ is the diameter of the cylinder, and $\nu$ is the kinematic viscosity. The simulations are performed for $Re=300$ and $f=0.1$. An overset grid with the dimension of $2D \times 2D$ discretized by $201$ grid points in both $x$ and $y$ directions is used around the cylinder. In addition, a square region at the center of the overset grid with the size of $1.6D \times 1.6D$ is blanked on the background grid. The background grid is fixed, however, the overset grid is fixed to the center of the cylinder and rotates with its motion. Dirichlet boundary condition is used for all outer boundaries of the background domain similar to~\citet{kim2006immersed} where the velocities are equal to zero on the boundaries, and the boundary condition for the overset grid are interpolated from the background grid using the Eq.~\ref{velocity} as
\begin{equation} 
u^{overset}=Q u_{interpolate}^{background}
\end{equation}
Figure~\ref{Fig:rotcylinder} shows the torque coefficient defined as $C_T=T/ (0.5 \rho_m U^2 D^2/2$) during the time, where $T$ is the torque and $\rho_m$ is the density of the fluid. In order to compare the numerical results, a simulation using a single grid with the same dimension as the background in overset simulation is performed in a non-inertial frame of reference. The single grid is discretized using $401$ grid point in both $x$ and $y$ directions which provides a grid resolution of $0.01 D$ near the cylinder. Fig.~\ref{Fig:rotcylinder} compares the results of the overset grid with a single grid for eight cycles. The result of torque coefficient using overset grid and single are in good agreement with each other and also they are in good agreement with the results of~\citet{borazjani2013parallel} and slightly lower than the result of ~\citet{kim2006immersed}.

\clearpage

\subsection{Forced inline oscillations of a cylinder in a fluid initially at rest}

\begin{figure}
	\centering
	\includegraphics[width=1.\textwidth]{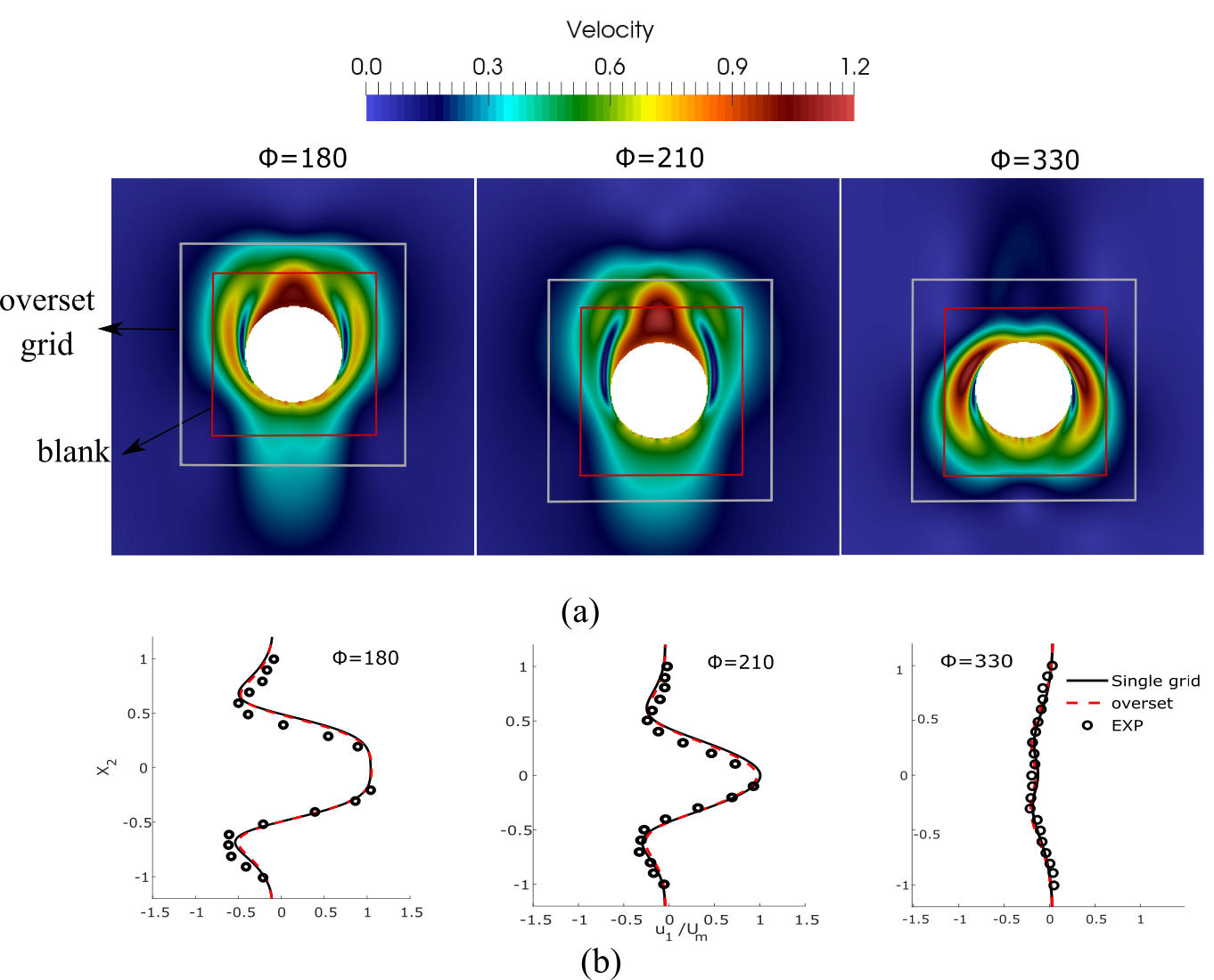}
	\caption{Results of numerical simulation for oscillatory cylinder a) contour of velocity for three different phase angles b) Comparison of the inline velocity component~$(u)$ profile at position $x_1=-0.6D$ for three different phase angles between numerical results~(overset: \textcolor{red}{- -}, single grid: $-$) and the experimental measurements~(\textbf{o}) of Dutsch et al.~\citet{dutsch1998low}.}
	\label{Fig:osicyl_velocity}
\end{figure}

The developed framework is validated for the case of a circular cylinder starting to oscillate in the horizontal direction in a fluid initially at rest. The translational motion of the cylinder is given by a harmonic oscillation:
\begin{equation} 
x_c(t) = - A_m sin(2\pi ft),
\end{equation}
where $x_c$ is the location of the center of the cylinder, $f$ is the oscillation frequency, and $A_m$ is the oscillation amplitude which result in two non-dimensional flow parameters, i.e., Reynolds number and Kuelegan-Carpenter number as follows:
\begin{equation} 
Re= \frac{U_m~D}{\nu}, ~~KC=\frac{U_m}{f~D}
\end{equation}
where $U_m$ is the maximum oscillation velocity, $D$ is the diameter of the cylinder, and $\nu$ is the kinematic viscosity of the fluid. The simulation is performed for $KC=5$ and $Re=100$, for which the experimental results have been reported by~\citet{dutsch1998low}. The  size of the background grid is $100 D \times 100 D$ which is discretized using $301 \times 301$ nodes, and $100 \times 100$ nodes are distributed uniformly in a $3D \times 3D$ box which contains the cylinder during the oscillations. $201 \times 201$ grid nodes are uniformly distributed in the smaller domain with the size of $2.4D \times 2.4D$ aligned and moving with the center of the cylinder~($x_c(t)$). In addition, a blank region with the size of $2.1D \times 2.1D$ is used to blank out the nodes in the background grid. The non-dimensional time-step of $\Delta t =0.0167$ is used for this simulation in both domains. The far-field boundary condition is applied to the boundaries of the background grid while the boundaries of the small grid are interpolated from the background domain. In addition, to compare the overset results with the results obtained using a single grid, a grid with the same dimension as the background grid explained above~($100D \times 100D$) is discretized using $401 \times 401$ grid nodes which provides a grid resolution of $0.01 D$ near the cylinder is used. The simulation for the single grid is performed in a non-inertial frame of reference. Fig.~\ref{Fig:osicyl_velocity}-a shows the position of the overset grid as well as the velocity contours for three the different angles in overset and background grids. The comparison between the inline velocity profiles at $x_1 = 0.6D$ for three different phase angles~($\phi=2 \pi f t$) calculated by our framework and experimental measurements by~\citet{dutsch1998low} is presented in Fig.~\ref{Fig:osicyl_velocity}-b. Our numerical results show good agreement with the experimental data.

\clearpage

\subsection{Freely falling circular cylinder}

 In this test case, we consider freely falling of circular cylinder under gravity using fluid-structure interaction in which the cylinder falls due to the gravitational and fluid forces. Assuming that the wake behind the cylinder is two dimensional and the cylinder is moving in an infinite fluid the numerical simulations are performed using a two-dimensional grid with symmetric boundary condition in two-dimensional direction~($z$) and far-field boundary conditions for all the boundary points in $x$ and $y$ direction. The acceleration of the body due to the gravitational and Buoyancy forces is $(\rho_s/\rho_f-1)g$, where $g$ is the gravitational acceleration, and $\rho_s$ and $\rho_f$ are the density of the cylinder and fluid, respectively. The Reynolds number is considered to be the same as the Galileo number defined as 
\begin{equation} 
Re= \frac{(|\rho_s/\rho_f-1|g)^{1/2} D^{3/2}}{\nu}
\end{equation}
where $({|(\rho_s/\rho_f-1)g D|})^{1/2}$ is the characteristic velocity, $\nu$ is the dynamic viscosity of water and $D$ is the diameter of the cylinder. Neglecting the body rotation, the equation of motion for the cylinder in the inertial frame of reference can be obtained using the two-dimensional Newton's equations of motion for a rigid body as
\begin{equation} 
M \frac{du}{dt}=F_f-(\rho_s-\rho_f) V g
\end{equation}
where $M=\rho_s \pi D^2/4$ is the mass of cylinder, $V$ is the cylinder's volume, and $F_f$ is the force exerted on the body by fluid in the non-inertial reference frame. The above equation can be written in non-dimensional form as 
\begin{equation} 
\frac{\pi}{4} \frac{\rho_s}{\rho_f} \frac{du^*}{dt^*}=F^*_f-\frac{\pi}{4}
\end{equation}
where $u^*$, $t^*$, and $F^*_f$  are the non-dimensional velocity, time, and fluid force, respectively. Considering density ratio of solid to fluid $\rho_s/\rho_f=2.5$, $\nu=8 \times 10^{-4} N.s/m^2$, and $D= 0.05m$ leads to $Re=53.61$ in this simulations. Both the fluid and the cylinder are initially at rest and the cylinder starts the free-fall abruptly after start of the simulations. Two test cases, one with a single cylinder~(section~\ref{Single cylinder-sec}) and the other with multiple cylinder~(section~\ref{Multiple cylinders-sec}), are performed for verifying our framework.

\subsubsection{Single cylinder} \label{Single cylinder-sec}
\begin{figure}
	\centering
	\includegraphics[width=1.\textwidth]{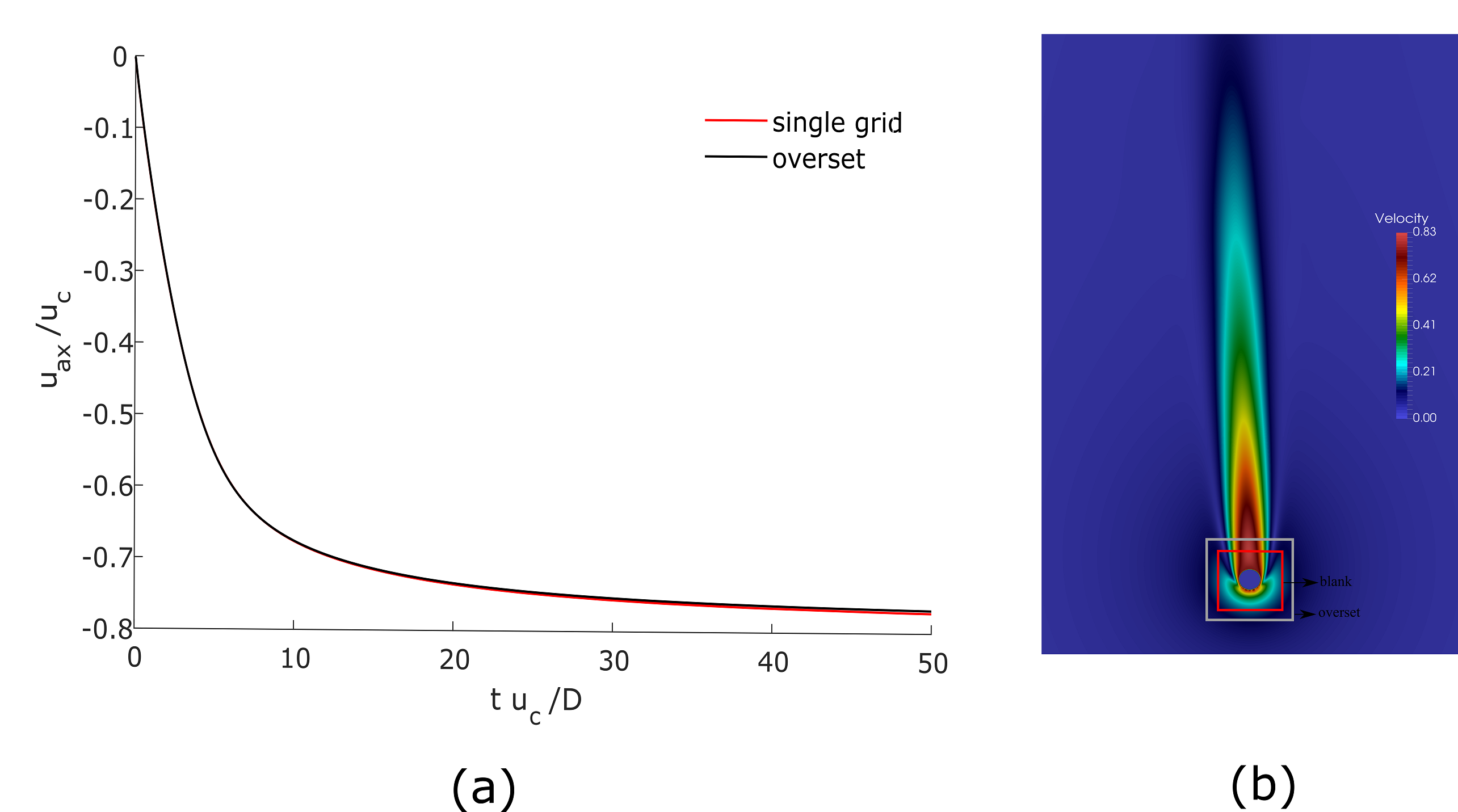}
	\caption{(a) Time histories of the velocity of the cylinder in the gravitational direction for the overset~($\textcolor{red}{-}$) and the single~($-$) grid simulations (b) contour of velocity magnitude at $t u_c/D=50$} 
	\label{Fig:single particle}
\end{figure}

A circular cylinder is placed in the domain similar to schematic setup presented in Fig.~\ref{Fig:schematic-cylinder}, for particle number $1$~(the bodies and overset grids for particle 2 and particle 3 are not included in this simulation). The background grid with the size of $60D$ in $x$ and $100D$ in $y$-direction is discretized using $301$ and $801$ grid points, respectively. The grid points for the background grid are distributed such that the spatial resolution is $0.06 D$ around the overset grid during the whole simulation. For the overset domain with the dimension of $4D \times 4D$, $201$ grid points are uniformly distributed in both $x$ and $y$ directions which provide the grid resolution of $0.02 D$. In addition, a blank region with the size of $3.4D \times 3.4D$ is used to blank out the nodes in the background grid. The flow in the overset grid is solved using a non-inertial frame of reference attached to the center of the cylinder which moves with the cylinder as it falls, whereas the equations for the background grid are solved in the inertial frame of reference. The simulation using overset grid framework is compared with the numerical results of a single grid with the same dimension as the background grid in the overset simulation, which was performed in a non-inertial frame of reference. For the single grid, $801$ and $1931$ grid points were distributed in $x$ and $y$ directions, respectively, which provides the spatial resolution of $0.02 D$ around the cylinder. Fig.~\ref{Fig:single particle} compares the time histories of the velocity of the cylinder in the gravitational direction for the overset grid and the single grid simulations. In both cases, the cylinder accelerates monotonically and almost reaches its terminal velocity around $t u_c/D=50$. As can be observed in Fig.~\ref{Fig:single particle} the results for the overset grid and the single grid are in good agreement with each other and the difference in the translational velocity of the cylinder using single grid and the overset grid at time $t u_c/D=50$ is around $0.2\%$.

\clearpage

\subsubsection{Multiple cylinders} \label{Multiple cylinders-sec}
\begin{figure}
	\centering
	\includegraphics[width=0.4\textwidth]{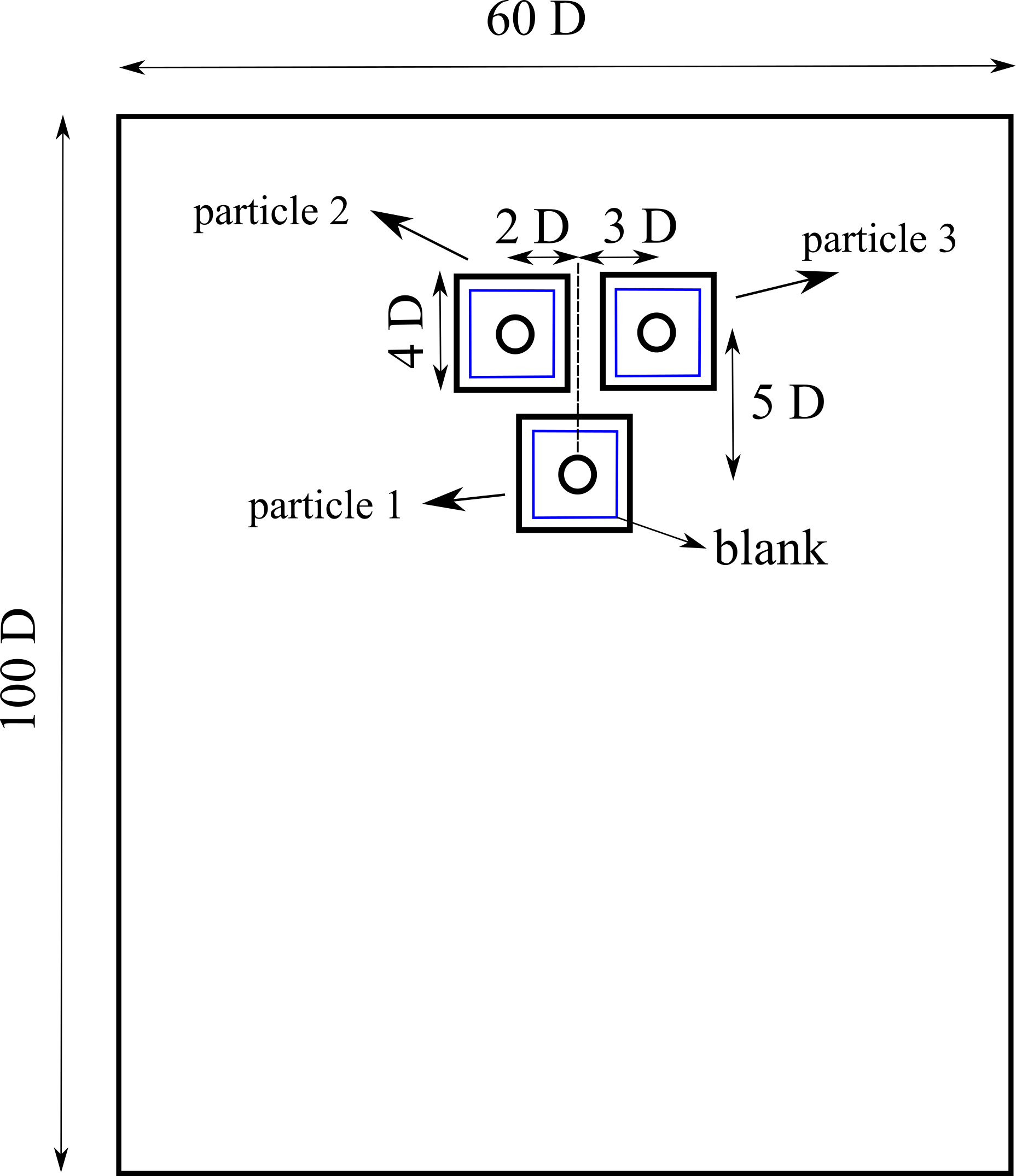}
	\caption{Schematic position of overset grids~(small black squares), background grid~(large rectangle domain), and blank region~(small blue squares) relative to each other for simulation of free fall of multiple circular cylinders under gravitational force.} 
	\label{Fig:schematic-cylinder}
\end{figure}

\begin{figure}
	\centering
	\includegraphics[width=1.\textwidth]{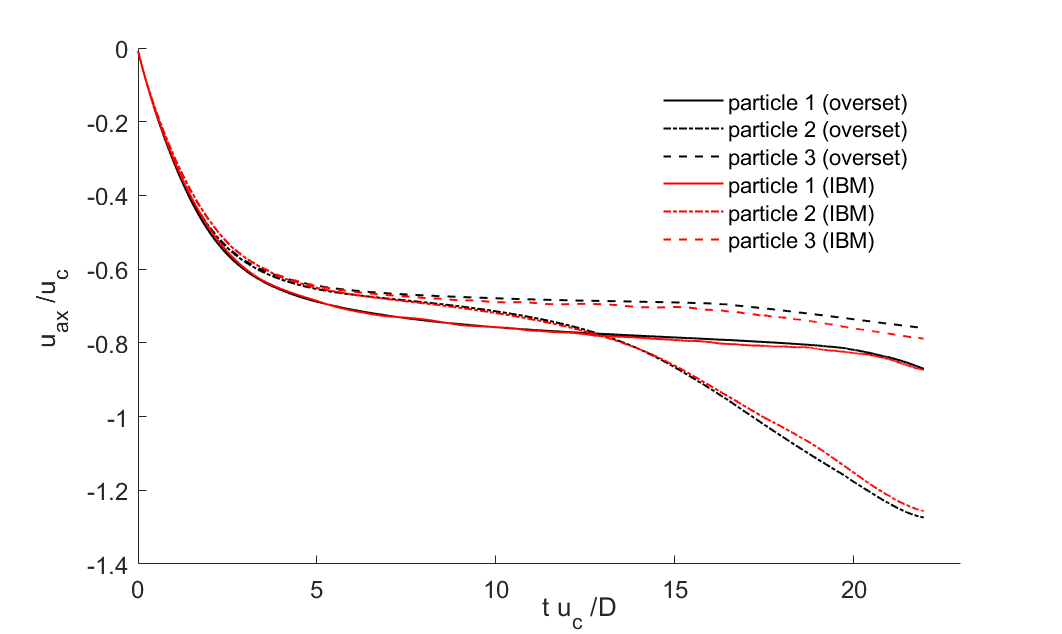}
	\caption{Comparison of time histories of the velocity for three different cylinder in the gravitational direction between overset and single grids.} 
	\label{Fig:multi-particle}
\end{figure}

\begin{figure}
	\centering
	\includegraphics[width=.8\textwidth]{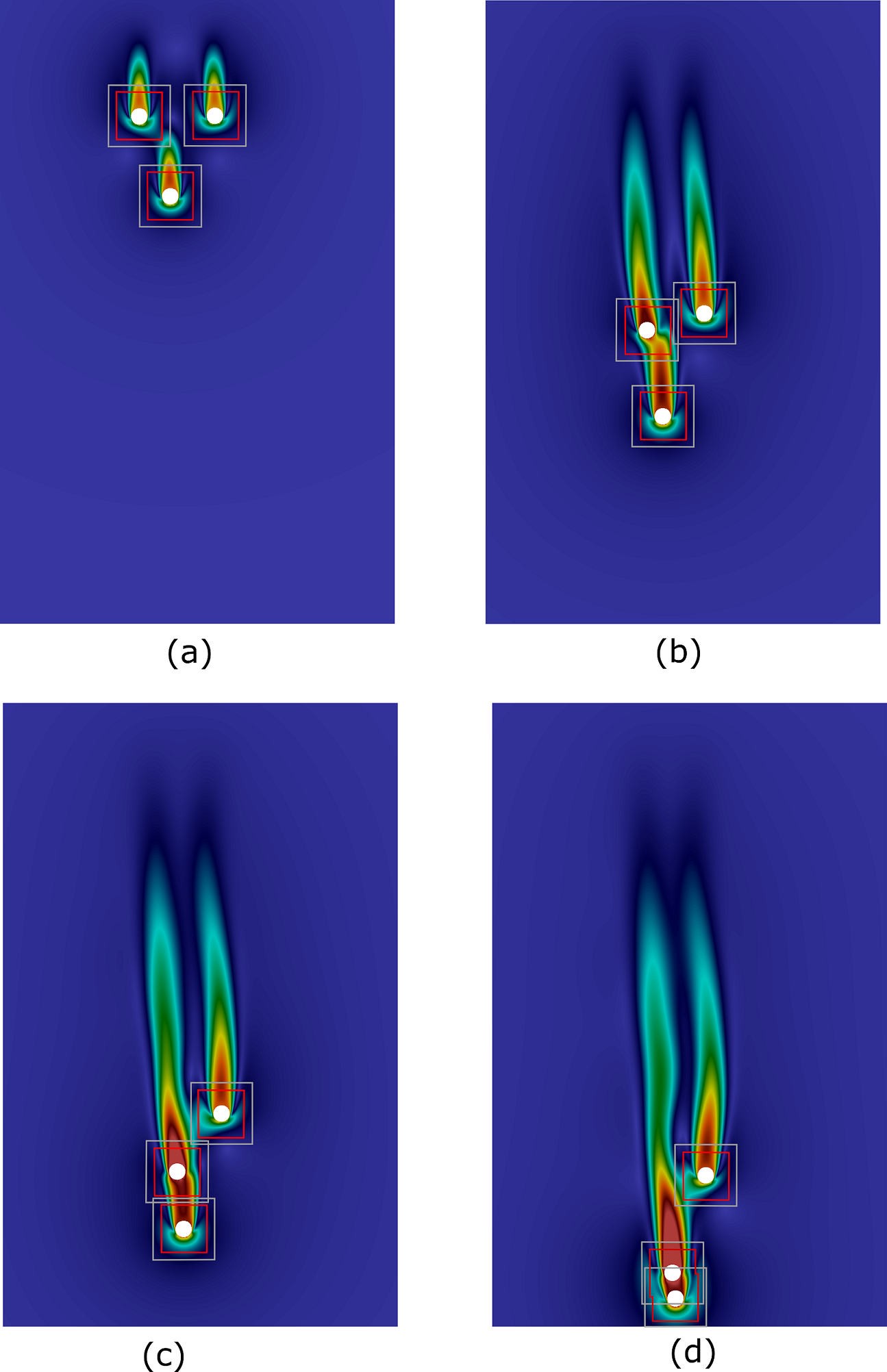}
	\caption{Contour of velocity magnitude for overset simulation of multiple circular cylinders at time a) $t u_c/D=4.4$ b) $t u_c/D=12$ c) $t u_c/D=15.6$ d) $t u_c/D=22$ during the simulation.} 
	\label{Fig:multi-particle_velocity}
\end{figure}

To test our framework for multiple overset grids, the simulation for free fall of multiple cylinders in an infinite flow is performed. Fig.~\ref{Fig:schematic-cylinder} shows the schematic setup used in this simulation. Three cylinders centered at $(0,0)$, $(3D,5D)$, and $(-2D,5D)$ are placed in the flow domain. An overset grid is generated around each cylinder with the dimension of $4 D \times 4D$ and the center is aligned with the center of each cylinder as showed in Fig.~\ref{Fig:schematic-cylinder}. The grid dimension and the number of grid points are the same as the overset simulation of single cylinder~(overset grids are discretized uniformly using $201$ grid points in each direction, and the background grid is discretized using $301$ and $801$ grid points distributed the same as the previous section). In order to verify the result of the overset grid the same simulation is performed using a single grid in an inertial frame of reference~(since a non-inertial frame of reference cannot be used for this simulation due to existence of multiple particles with different velocities). The computational domain is discretized the same as the single grid used in previous section~($801$ and $1931$ grid points are distributed in the $60 D \times 100 D$ domain which guarantees the spatial resolution of $0.02 D$ through the trajectory of the cylinder at all times). The simulations are performed using a strong-coupling fluid-structure interaction for both overset and single grid cases \citep{borazjani2008curvilinear}. Figure~\ref{Fig:multi-particle} compares the time history of translational velocity of cylinders in the gravitational direction for a single grid versus overset grids. The results of the overset grids and single grid are in good agreement with each other considering that different method and grid are used for each simulation. The maximum difference in translational velocity of cylinders using a single grid compared to the overset grid is observed for cylinder number 3 which is around $4\%$. Fig.~\ref{Fig:multi-particle_velocity} shows the position of the overset grids relative to each other as well as the background grid at several time instants. In addition, the special scenarios which discussed in section~\ref{Handling_special_cases} regarding the overlap of overset boundaries with immersed boundaries or other overset grids can be observed in this figure.


\clearpage

\subsection{Multiple mackerels in diamond arrangement}

To show the capability of our numerical framework in handling complex biological flows, a simulation with multiple self-propelled mackerels swimming in the diamond arrangement is performed. The swimmers can move in the background domain, thus, a high-resolution grid would be required in the path of the swimmers. While using a single grid can drastically increase the computations due to a huge number of grid points required, the overset method can provide high-resolution grid locally around the swimmers without considerably increasing the total number of grid points. 

The geometry of the mackerels used in this study is exactly the same as the previous simulations by~\citet{borazjani2010role} and \citet{borazjani2013parallel}. The kinematic motion of the mackerels is approximated by a backward traveling wave with the largest wave amplitude at the fish tail. The lateral undulations of the swimmers' body in non-dimensional form~(all lengths are non-dimensionalized with the fish length $L$) can be described as 
\begin{equation} \label{fish-motion}
h(z,t)=a(z) sin(2 \pi z / \lambda-2 \pi f t)
\end{equation}
where $z$ is the axial direction measured along the fish axis from the tip of the fish head; $h(z,t)$ is the lateral excursion of the body at time $t$; $a(z)$ is the amplitude envelope of lateral motion as a function of $z$; $\lambda$ is the wavelength, and $f$ is the frequency of the backward traveling wave. The amplitude envelop for a typical mackerel can be approximated by a quadratic curve~\citep{borazjani2008numerical}
\begin{equation} 
a(z)=a_0+a_1 z+a_2 z^2
\end{equation}
where $a_0$, $a_1$, and $a_2$ are chosen to be $0.02$, $0.08$, and $0.16$, respectively, to match the experimental curve of~\citet{videler1984fast} obtained for a typical mackerel. The maximum displacement of mackerel occurs at its tail $h_{max}= 0.1L$. The non-dimensional wavelength is chosen to be $\lambda/L= 0.95$ based on the experimental data by~\citet{videler1984fast}. The simulations are discretized using $240$ time steps per a tail beat period, which corresponds to a non-dimensional time step of $\Delta t= 1.39 \times 10^{-3}$. The Strouhal number~$(St)=fL/U$ and Reynolds number are chosen to be $0.6$, and $4000$, respectively, which has been shown to result in the final non-dimensional average velocity~$(Ut/L)$ close to $1$ during a a self-propelled steady-state simulation of a mackerel~\citep{borazjani2008numerical}.

The side swimmers are placed $0.45 L$ and $1.45 L$ laterally and posteriorly, respectively, relative to the front swimmer~(where $L$ is the fish length) and the last swimmer is placed $2.9 L$ behind the front swimmer. The background grid is a cuboid with dimensions of $4.2 L \times L \times 14L$ (in, x,y and z directions, respectively), and the overset grids are also cuboids with dimensions of $0.8 L \times 0.5 L \times 1.5 L$. Each fish is placed at the center of its corresponding overset grid. A region with dimensions of $0.6 L \times 0.3 L \times 1.3 L$ inside each overset grid is blanked from the background grid, whose solution is interpolated from the inner overset grids. The background grid is discretized by $9.7$ million grid nodes using a uniform mesh with constant spacing $\Delta x= 0.0187 L$. Each overset grid is discretized with a uniform mesh with spacing $\Delta x= 0.005 L$ in all directions with $161 \times 101 \times 301$ nodes results in $4.9$ million grid nodes. Therefore, the total number of grid points in this simulation is about $29$ million while a single grid with a similar resolution would require at least $470$ million grid points which is impractical for strong-coupling FSI simulations.

The Naiver-stokes equations are solved in a non-inertial frame of reference for moving overset grids where the reference frame is attached to the center of mass of the fish, however, the background grid is solved in an inertial frame of reference. Slip wall boundary condition is applied on the boundaries of the background grid and the boundaries of overset grids are interpolated from the background grid. The body motion of fish relative to the center of mass is prescribed as mentioned in Eq.~\ref{fish-motion} and there is no phase difference between the backward traveling waves of different swimmers. The velocity of the frame for each grid~(center of mass for each fish) was calculated based on the fluid forces on the body of the fish with two degree of freedom, in $x$ and $z$ directions, using a strongly-coupled fluid-structure interaction strategy~\citep{borazjani2008curvilinear}.

\begin{figure}
	\centering
	\includegraphics[width=0.9\textwidth]{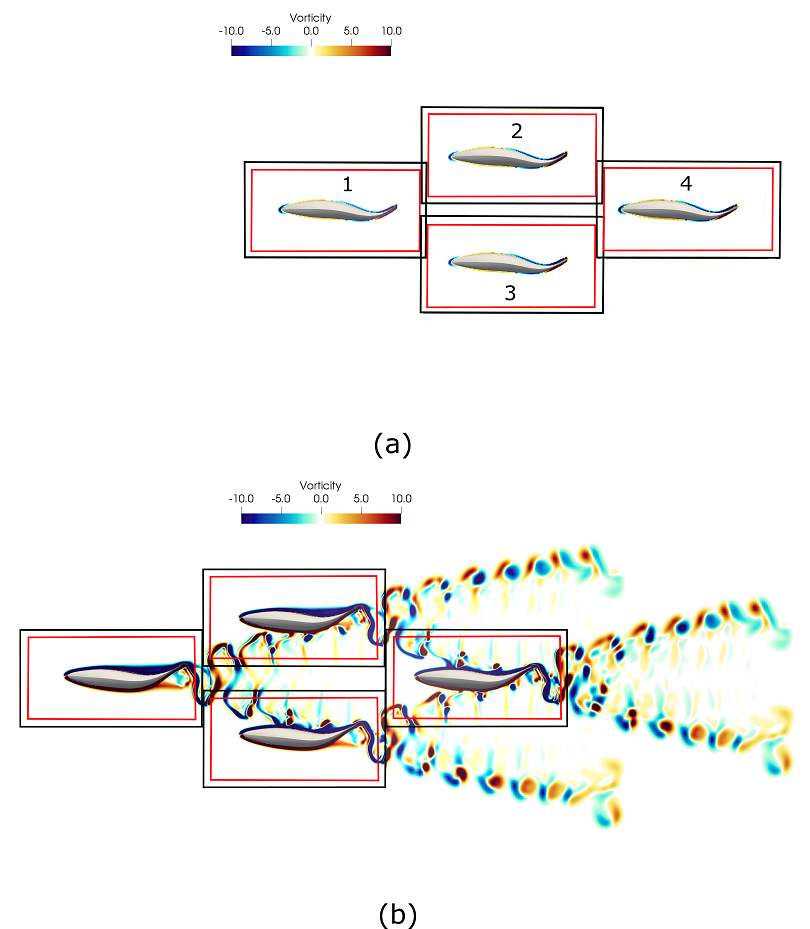}
	\caption{Contours of vorticity in the midplane of the four fish swimming in the diamond arrangements. a) initial position of overset grids b) position of overset grids at $t/T= 8.8$  .contour of each overset grid and background grid are consistent. vortical structures are are advected from one domain to the other. Thick black lines represent the boundaries of the overset grids and thick red lines shows the blank region.} 
	\label{Fig:fish_vorticity}
\end{figure}

Figure~\ref{Fig:fish_vorticity} shows the out of plan vorticity contours on the midplane of the fish as well as the position of overset grids compared to each other and background grid. As it can be seen in this figure the solution is consistent over the overset grids and the background grid and the vortical structures are advectedand from one domain to the other. It can be observed that the wake of each mackerel bifurcates into two rows of vortices~(double row structure), which is expected for the Strouhal number of 0.6 based on the work by~\citet{borazjani2008numerical}, and interact with downstream immersed bodies and their wakes. Fig.~\ref{Fig:fish_vorticity} and Fig.~\ref{Fig:multi-particle_velocity} show that our parallel framework~(regardless of parallel grid distribution) is capable of handling different scenarios regarding the relative position of overset grids and blank regions compared to each other including multiple grid overlapping, overset interface intersection with interface/blank as well as immersed bodies. 

\begin{figure}
	\centering
	\includegraphics[width=1\textwidth]{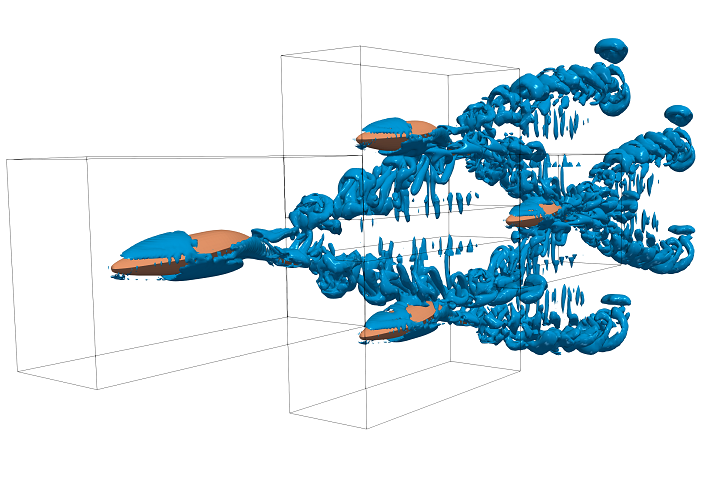}
	\caption{The 3D vortical structures visualized by the iso-surfaces of Q-criteria for four fish swimming in diamond arrangement.} 
	\label{Fig:fish_Q}
\end{figure}

\begin{figure}
	\centering
	\includegraphics[width=1\textwidth]{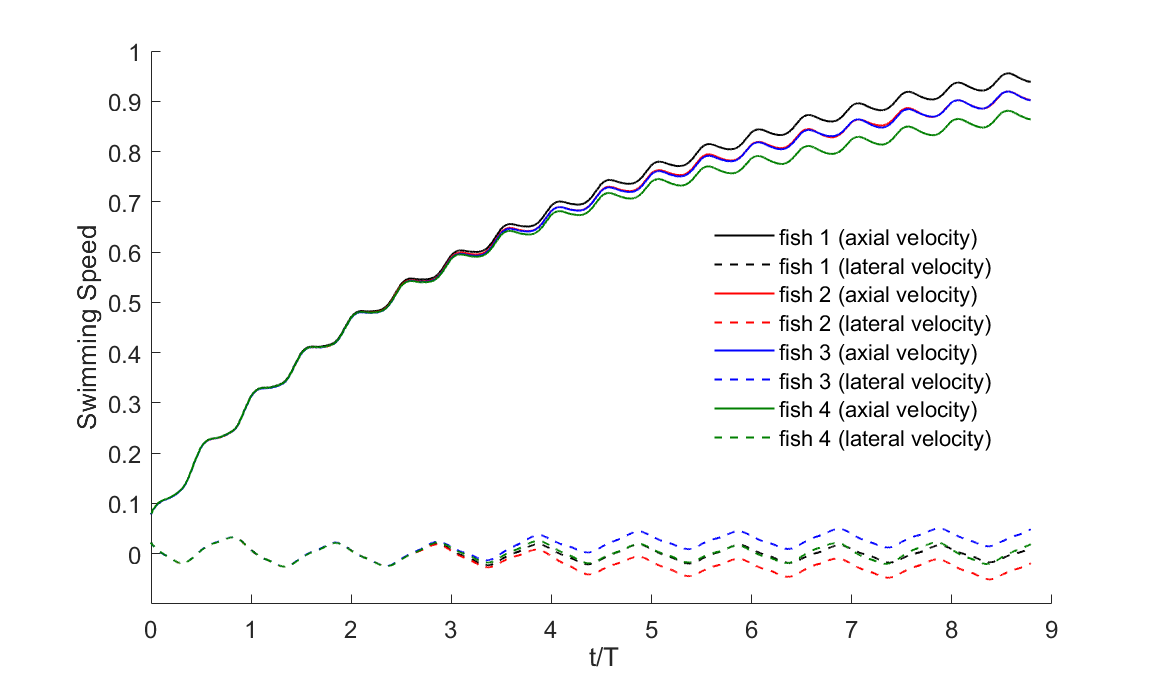}
	\caption{Comparison of the self-propelled fish swimming axial and lateral speed for different swimmers in diamond shape. Fish $1$ to $4$ are denoted in Fig.~\ref{Fig:fish_vorticity}} 
	\label{Fig:fish_velocity}
\end{figure}

Figure~\ref{Fig:fish_Q} shows the the 3D flow field  visualization using iso-surface of Q-criteria generated by the swimmers. The wake of each swimmer bifurcates into two rows of vertices and interact with the wake of downstream fish. The swimmers can move relative to each other with two degrees of freedom in the lateral and axial directions and thus they can have different velocities in these directions. Fig.~\ref{Fig:fish_velocity} compares the axial and lateral velocity for all the swimmers during the time. As can be observed the leading swimmer has the highest axial velocity among all while the last swimmer has the lowest one. However, swimmers on the side~(swimmers $2$ and $3$) have a higher lateral velocity compared to the swimmer 1 and 4 which have almost the same lateral velocities.

\clearpage

\section{Parallel efficiency}\label{Parallel-efficiency}
In this section, the swimming simulation is used to investigate the speedup for different parts of our solver. The total number of grid points in this simulation is about $30$ million grid points and approximately $1.2$ million query points. The simulations are run using $560$ core on Terra cluster at Texas A$\&$M University, which contains $320$ computing nodes, each node contains $2$ Intel Xeon E5-2680 v4 2.40GHz 14-core each, and uses Intel Omni-Path as the cluster-interconnect. Table~\ref{time} represent the wall-clock time for the CURVIB flow solver, grid assembly task, and interpolation for up to $560$ cores. As can be seen from this table, the computational time required for the grid assembly is relatively very small compared to the computational time needed for the flow solver. The grid assembly time decreases for up to $140$ cores while after that the time does not show a significant change. However, even using $560$ the time required for the grid assembly is about $7\%$ of the flow solver. In addition, comparing the interpolation time and the flow solver time shows the efficiency of our interpolation method, where using all number of cores, it is less than $0.1\%$ of the flow solver time.

\begin{center}
\begin{tabular}{ |p{3cm}||p{3cm}|p{3cm}|p{3cm}|  }
 \hline
 \multicolumn{4}{|c|}{Wall-clock time~(sec)} \\
 \hline
 No. of cores & flow solver & grid assembly & interpolation\\
 \hline
 $8$   & $440.3$ & $5.1$    & $1.1 \times 10^{-1}$\\
 $84$  & $41.5$  & $0.54$ & $2.7 \times 10^{-2}$\\
 $112$ & $32.3$  & $0.45$ & $2.1 \times 10^{-2}$\\
 $140$ & $28.1$  & $0.40$ & $1.8 \times 10^{-2}$\\
 $280$ & $14.4$  & $0.43$ & $9.8 \times 10^{-3}$\\
 $420$ & $11.2$  & $0.58$ & $7.6 \times 10^{-3}$\\
 $560$ & $8.2$   & $0.58$ & $5.1 \times 10^{-3}$\\
 \hline
\end{tabular}\label{time}
\end{center}

\begin{figure}
	\centering
	\includegraphics[width=.7\textwidth]{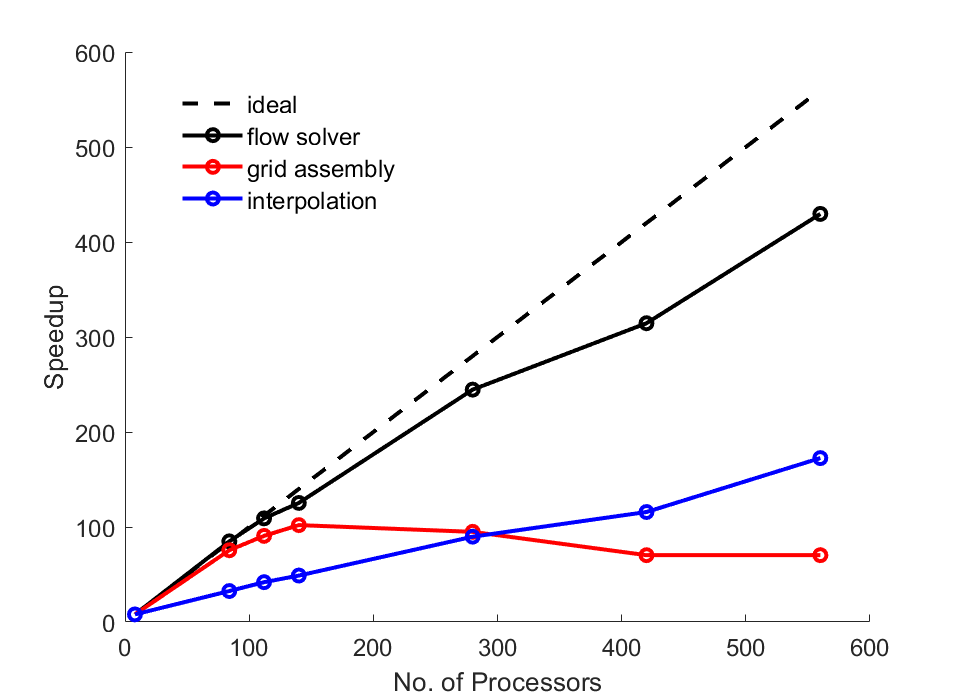}
	\caption{Wall-clock time speedup versus the number of processors for the fish schooling using $560$ processors.} 
	\label{Fig:speedup}
\end{figure}

Figure~\ref{Fig:speedup} shows the strong scalability for wall-clock time for different part of our overset-CURVIB solver. As can be observed our flow solver shows good speedup for the maximum number of processors used in this work~($560$ processors). The grid assembly kernel has a speedup close to ideal for up to $140$ processors while after that the scalability drops. The reason for speedup drop-off using more than $140$ processors is the load-imbalance for grid assembly method using our available grid partitioning strategy in which all the grids are distributed to all the available processors. As previously mentioned, this partitioning strategy results in the best speedup for the flow solver while can increase the communication cost and overhead for the grid assembly. However, by comparing the time required for flow solver to the time of the grid assembly kernel~(Table~\ref{time}), optimizing the grid partitioning for flow solver is more reasonable. Using another partitioning strategy for load balancing, depending on the problem, in future to balance the number of query points in each processor can help to improve the speedup for more processors. However, the load balancing is out of the scope of this work. Finally, the speedup for interpolation is also presented in Fig.~\ref{Fig:speedup}. The interpolation's speedup is not close to ideal, however, considering the small computational time required to interpolation this speedup is not unexpected. 

\section{Conclusions} \label{Conclusions}
We developed a new parallel dynamic overset-CURVIB framework by extending the previous overset-CURVIB method~\citep{borazjani2013parallel} for fixed overset grids and a sequential grid assembly to moving overset grids with an efficient parallel grid assembly. Our new framework utilizes a non-inertial frame of reference to solve the moving/rotating overset grids to avoid recalculating the curvilinear metrics of transformation while the background/stationary grids are solved in the inertial frame. In addition, a sharp-interface curvilinear immersed boundary method as well as an strong-coupling FSI method are used to handle solid immersed bodies in the domain in the context of our CURVIB flow solver. The framework enables us to perform high-resolution fluid-structure interaction simulations of real-life complex flows, which could not be handled with our previous strategy. Using dynamic overset grids allows us to increase the grid resolution locally around moving immersed bodies without drastically increasing the total number of grid points in simulations. 

Major developments of this work compared to the previous method~\citep{borazjani2013parallel} are: 1) developing a new grid assembly algorithm for partitioned grids~(parallel distributed environment); 2) using a new walking strategy for donor search; 3) developing a new algorithm for variable interpolation by forming an interpolation matrix; 4) directly integrating the grid assembly kernel into the flow solver instead of using an out-of-core strategy; 5) extending our previous framework to handle moving overset grids in a non-inertial frame of reference while stationary ones in an inertial frame.  

The major challenge in developing a parallel dynamic overset framework is the need for an efficient parallel communication strategies to transfer information between subdomains for a domain decomposition in which all grids are distributed to all processors~(optimal domain decomposition for our flow solver). Several steps have been made to increase the scalability and decrease the computational/communication cost of our framework including: 1) using OBBs to decrease the search space; 2) using the control cells to accelerate the donor search; 3) data packing to combine multiple messages into a single message which results in decreasing the total number of communications and consequently decreases the overhead associated with it; 4) using non-blocking data transfer to reduce the overhead and maximize the communication/computation overlap; and 5) developing a vectorized implementation for data interpolation in parallel which can drastically decrease the interpolation time. The parallel scalability of our solver is tested for different part of our framework for the school of swimmers test case. While a good scalability is achieved for our flow solver for up to $560$ processors, the scalability of grid assembly kernel drops off for more than $140$ processors due to the load-imbalance related to the partitioning strategy used in this work~(as discussed previously in section~\ref{Parallel-efficiency}). A better initial partitioning strategy that takes the communications costs of the overset grids into account in future can help to enhance the scalability of grid assembly kernel. Nevertheless, the time required for the grid assembly is less than 7\% of the total simulation time even at the highest number of CPUs tested (560 cores). 

Finally, the new framework is verified and validated against experimental data, the analytical solution of Taylor-Green vortex, and other benchmark solutions and the capability of the new framework is shown by performing multiple circular cylinders in a free fall under gravity in a fluid domain as well as school of swimmers in a diamond shape. Using overset grids reduced the total number grid points from $500$ to $30$ million while preserving the same resolution in the self-propelled fish school. This new framework enables us to tackle challenging real-world problems which cannot be handled without moving overset grids. 

\section*{Acknowledgement}
This work was supported by American Heart Association~[AHA: Grant No. 13SDG17220022], National Science Foundation~[NSF: CAREER grant CBET 1453982], and the computational resources were provided by the Texas A$\&$M High Performance Research Computing Center~(HPRC).

\bibliographystyle{cas-model2-names}

\bibliography{cas-refs}


\end{document}